\documentclass[11pt,a4paper]{article}
\usepackage{amssymb}
\usepackage{graphicx}
\usepackage{amsmath}
\usepackage{makeidx}
\usepackage{indentfirst}
\usepackage[T1]{fontenc}
\usepackage[utf8]{inputenc}
\usepackage{authblk}

\newcounter{resultnum}[section]
\newcounter{conclusionnum}[section]
\newcounter{conditionnum}[section]
\newcounter{conjecturenum}[section]
\newcounter{examplenum}[section]
\newcounter{exercisenum}[section]
\newcounter{lemmanum}[section]
\newcounter{notationnum}[section]
\newcounter{theoremnum}[section]
\newcounter{definitionnum}[section]
\newcounter{corollarynum}[section]
\newcounter{remarknum}[section]
\newcounter{propositionnum}[section]
\newcounter{acknowledgementnum}[section]
\newcounter{algorithmnum}[section]
\newcounter{axiomnum}[section]
\newcounter{casenum}[section]
\newcounter{claimnum}[section]
\newcounter{summarynum}[section]
\newcounter{problemnum}[section]
\begin{document}

\title{Modified Einstein and Finsler Like Theories on Tangent Lorentz Bundles%
}
\date{September 18, 2014}
\author{\textbf{Panayiotis Stavrinos}\thanks{%
pstavrin@math.uoa.gr}  \and  \textsl{\small Department of Mathematics,
University of Athens } \\
\textsl{\small Ilissia, Athens, 15784 Greece } \\
{\qquad} \\
\textbf{Olivia Vacaru}\thanks{%
olivia.vacaru@yahoo.com} \\
{\small {\textsl{National College of Ia\c si;\ 4 Arcu street, Ia\c si,\
Romania,\ 700125 }}} \\
{\qquad} \\
\textbf{Sergiu I. Vacaru}\thanks{%
sergiu.vacaru@cern.ch; sergiu.vacaru@uaic.ro} \\
{\small {\textsl{Theory Division, CERN, CH-1211, Geneva 23, Switzerland;}}}%
\thanks{%
visiting researcher}\\
{\small \ and {\textsl{\, University "Al. I. Cuza" Ia\c si}, Rector's
Office;\ } }\\
{\small {\textsl{\ 14 Alexadnru Lapu\c sneanu street, Corpus R, UAIC, office
323,} }}\\
{\small {\textsl{I}a\c si, Romania 700057}} }
\maketitle

\begin{abstract}
We study modifications of general relativity, GR, with nonlinear dispersion
relations which can be geometrized on tangent Lorentz bundles. Such modified
gravity theories, MGTs, can be modeled by gravitational Lagrange density
functionals $f(\mathbf{R},\mathbf{T},F)$ with generalized/ modified scalar
curvature $\mathbf{R}$, trace of matter field tensors $\mathbf{T}$ and
modified Finsler like generating function $F$. In particular, there are
defined extensions of GR with extra dimensional "velocity/ momentum"
coordinates. For four dimensional models, we prove that it is possible to
decouple and integrate in very general forms the gravitational fields for $f(%
\mathbf{R},\mathbf{T},F)$--modified gravity using nonholonomic 2+2 splitting
and nonholonomic Finsler like variables $F$. We study the modified motion
and Newtonian limits of massive test particles on nonlinear geodesics
approximated with effective extra forces orthogonal to the four--velocity.
We compute the constraints on the magnitude of extra--accelerations and
analyze perihelion effects and possible cosmological implications of such
theories. We also derive the extended Raychaudhuri equation in the framework
of a tangent Lorentz bundle. Finally, we speculate on effective modelling of
modified theories by generic off--diagonal configurations in Einstein and/or
MGTs and Finsler gravity. We provide some examples for modified stationary
(black) ellipsoid configurations and locally anisotropic solitonic
backgrounds.

\vskip0.1cm


\vskip3pt


\end{abstract}



\section{Introduction}

The late--time cosmic accelerating discovered and confirmed in 1998-1999
\cite{riess,perlmutter} opened new directions of research in cosmology and
seems to change paradigms in modern gravity and standard particle physics.
In spite of various efforts, the source of acceleration of Universe, dark
energy and dark matter effects etc are far from being understood. For
reviews of theoretical works and observational data, see \cite%
{peebls,padmanabhan,nojiri,tsujikawa,harko} and references therein.
Different theoretical models in which the Einstein--Hilbert action is
replaced by functions $f(R,q),$ (where $R$ is the Ricci scalar and a
function $q$ is be used, for instance, for the trace of energy--momentum of
matter, torsion fields etc), have been investigated in a number of papers.
These have stated the conditions of existence of viable cosmological models,
analysed the constraints obtained from the classical tests of general
relativity and quantum gravity models, studied the galactic dynamics and
test particle propagations with and without dark matter etc. They have also
explored possible connections with Modified Newtonian Dynamics (MOND) and
the Pioneer anomaly and considered the astrophysical and cosmological
implications of non--minimal coupling matter--geometry models \cite%
{moffat,bertolami,thakur,sami,harko1}, Finsler like generalizations \cite%
{stavrinos,stavrinosk,mavromatos,sindoni,visser,vcrit} etc.

The gravitational field equations in general relativity, GR, and
extra--dimensional extensions (including models on (co) tangent bundles with
commutative and noncommutative variables) have been found to possess a
decoupling property with respect to certain nonholonomic frames of
reference. This allows us to integrate such systems of partial differential
equations (PDE) in general off--diagonal forms \cite{veyms,vfbr}.\footnote{%
The metrics for these classes of solutions can not be diagonalized via
coordinate transform and the geometric/ physical objects may depend on all
coordinates via generating and integration functions and various parameters.}
\ Such methods of constructing off--diagonal solutions in various gravity
theories were elaborated by introducing Finsler like variables in Einstein
gravity and various modifications.

Finsler like variables can be naturally introduced on a (co) tangent bundle,
$T\mathbf{V}$, to a Lorentz manifold, $\mathbf{V}$, with possible
noncommutative extensions, for various classical and quantum gravity models
with modified dispersion relations \cite{vmdrhl,vfbr}. Considering
non--integrable (equivalently nonholonomic/anholonomic) 2+2 splitting on
(pseudo) Riemannian spacetimes, we can mimic certain locally anisotropic
configurations with prescribed fibred local structures. A Finsler nonlinear
quadratic element $F(x,y)=F(x^{k},y^{a}=dx^{a}/d\tau )$ is present in such
theories as a nonlinear generating function/metric for three fundamental
geometric objects. These are the nonlinear connection (N--connection), $%
\mathbf{N}=\{N_{i}^{a}(x,y)\}, $ a lift to total metric, $\mathbf{g}%
=\{g_{\alpha \beta }\},$ and distinguished connection, $\mathbf{D}=\{\mathbf{%
\Gamma }_{\beta \gamma }^{\alpha }\}$ (d--connection; which is different
from the Levi--Civita connection, $\nabla =\{\Gamma _{\beta \gamma }^{\alpha
}\}$).\footnote{\label{foot2} We write boldface symbols for spaces (and
geometric objects on such spaces) enabled with so--called horizontal (h) and
vertical (v) splitting, $\mathbf{N}:\ T\mathbf{V}=h\mathbf{V}\oplus v\mathbf{%
V,}$ when coordinates and indices split in the form $u^{\alpha
}=(x^{i},y^{a})$ [in brief, $u=(x,y)$] and $\beta =(j,b).$ Such
h-v--splitting exist naturally of vector/tangent bundles, but can be
introduced formally, for instance, as a $2+2$ splitting to diads on (pseudo)
Riemannian manifolds (when $i,j...,=1,2$ and $a,b,...=3,4$). For (pseudo)
Finsler models on tangent bundles, the indices run values of type $%
i,j...,=1,2,3,4$ and $a,b,...=6,7,8,9.$ The term "pseudo" will be used for
any necessary local signature of metrics $(\pm, \pm, \pm, \pm )$. Readers
may consider details on such constructions in Refs. \cite{vrflg,veyms}.}
There are various models of (generalized) Finsler geometry and gravity
theories depending on explicit assumptions on data $\left( F:\mathbf{g,N,D}%
\right) $ and how corresponding curvature, torsion and other tensors are
derived. Such theories can be metric compatible, or noncompatible, see
critical remarks in \cite{vcrit,vrflg}. In all cases, minimal extensions of
GR with a well defined axiomatic can be encoded into certain extended
principles of general covariance and relativity. Considering sections $%
y^{a}(x^{k})$ on a basic Lorentz manifold, we generate certain effective
"osculating" (pseudo) Riemannian metrics $\tilde{g}_{ij}(x):=\mathbf{g}%
_{ij}(x,y(x))$ derived in nonlinear forms from $F(x,y(x)) $ (such approaches
were considered in Finsler modifications of gravity and cosmology in \cite%
{stavrinos,stavrinosk}). This class of theories is with $f(R,T,...)$
modifications of GR and (via nonholonomic frame and off--diagonal
deformations of metrics and distortions of fundamental geometric structures)
can be related to various anisotropic modifications of Ho\v{r}ava--Lifshitz,
and/or covariant anisotropic models of gravity etc \cite%
{nojiri,tsujikawa,harko,harko1,vmdrhl}. Conventionally, we denote such
modified gravity theories with gravitational Lagrangians $f(R,T,F)$.

It is the purpose of the present article to study extensions of standard GR
to certain forms with Lagrange density $f(R,T,F)$ when the constructions can
be modelled by generic off--diagonal and nonholonomic effects in an
effective Einstein or Finsler like gravity theory. We shall consider
standard models of matter with respect to certain N--adapted frames of
reference which via nonholonomic constraints and off--diagonal interactions
may mimic exotic fluids and states of matter, quantum effects. Such theories
are with modified dispersion relations and/or anomalies, anisotropies and
"non--compactified" extra dimensions with (co) velocity like variables etc.
The field equations and the covariant divergence of the stress--energy
tensor can be derived in two equivalent geometric and variational forms
working with N--adapted geometric objects. Nonholonomic constraints and
off--diagonal gravitational interactions can model nontrivial matter
configurations, for instance, nonlinear scalar field interactions. So, there
are possible alternative explanations for inflation scenarios, late time
accelerations and dark energy/matter effects. We shall demonstrate the
possibility of reconstruction of various types Friedman-Lema\^
itre-Robertson-Worker (FLRW) cosmology and anisotropic modifications by
appropriate choices of the above mentioned functionals and/or
generating/integration functions for generic off--diagonal solutions.

We shall speculate on possible theories of reduction tangent bundle models
to standard ones with effective Einstein equations in GR, for $%
F(x,y)\rightarrow F(x,y(x)).$ In general, such constructions may result in
nontrivial torsion, non-zero covariant divergence of the stress--energy
tensor etc. We argue that following certain general principles on metric
compatible constructions completely determined by a fundamental metric
tensor we can provide an equivalent encoding of off--diagonal coefficients
of metrics into nonholonomic frames. We will state the conditions so that $%
f(R,T,F)$ with general data $\left( F:\mathbf{g,N,D}\right) $ can be
effectively described by certain $\left( \tilde{g}_{ij}(x),\nabla (x)\right)
$ and/or equivalent $(\mathbf{g,D=\nabla +Q)}$; all geometric objects being
determined by the same metric structure. The motion of massive test
particles in such modified theories is modeled by nonlinear geodesic
configurations (with effective extra acceleration). This is due to the
off--diagonal/nonholonomic interactions and nonlinear coupling between
matter and geometry. We shall investigate the Newtonian limit of such models
and compute certain expressions for the extra--acceleration. The
observational data for the perihelion of the Mercury can be used to impose a
general constraint on magnitude of such extra--acceleration and local
anisotropy effects.

The present paper is structured as follows. Section \ref{sfg} is devoted to
a brief introduction into the Finsler osculating gravity and its relation to
GR and modifications. There are derived field equations of $f(R,T,F)$
gravity. Some particular cases and the conditions of effective modeling via
generic off--diagonal solutions in GR and Finsler--Cartan gravity are
analysed in section \ref{spc}. We briefly discuss the procedure of
reconstructing gravity theories with scalar field and off--diagonal
interactions. The equations of constrained motion in modified backgrounds of
massive test particles, and the corresponding Newtonian limits of effective
locally anisotropic models, are analyzed. In section \ref{sdif}, we develop
a geometric method of decoupling and integrating the field equations in
modified gravity. We show how such equations can be solved in generic
off--diagonal form as nonholonomic deformations of de Sitter black holes to
certain rotoid and/or locally anisotropic solitonic configurations. Finally,
we discus and conclude our results in section \ref{sdc}.

\section{Modified Einstein \& Finsler Osculating Gravity}

\label{sfg}In the present section, we provide an introduction into modified
theories with local anisotropies which can be modelled as effective GR
theories for certain nonholonomic constraints resulting in zero torsion
structure but generic off--diagonal terms in metrics.

\subsection{Motivations for the $f(R,T,F)$ gravity}

We analyze two approaches to modifications of the GR theory.

\subsubsection{Action for $f(R,T)$ theories}

We can consider models on a four dimensional (4--d) pseudo--Riemannian
manifold enabled with metric structure $g_{ij}(x^{k})$ defining a
quad\-ratic linear element%
\begin{equation}
ds^{2}=g_{ij}(x)dx^{i}dx^{j},  \label{qel}
\end{equation}%
for $x=\{x^{k}\},$ a gravity theory corresponding to action%
\begin{equation*}
S=\int \sqrt{|g|}d^{4}x\{(16\pi )^{-1}f(R,T)+\ ^{m}L\},
\end{equation*}%
where $R=g^{ij}R_{ij}$ is the scalar corresponding to contraction of the
inverse metric $g^{ij}$ with the Ricci tensor $R_{ij}$ constructed for the
Levi--Civita connection $,$ $^{m}L$ is the matter Lagrangian density which
via corresponding variational calculus results in the stress--energy tensor,
\begin{equation*}
T_{ij}=-2(\sqrt{|g|})^{-1}\delta (\sqrt{|g|}\ ^{m}L)/\delta g^{ij},
\end{equation*}
and its trace, $T=g^{ij}T_{ij}.$\footnote{%
We use the natural system of units when the Newton constant, $G,$ and light
speed, $c,$ are subjected to the conditions $G=c=1$ and the gravitational
constant is $\kappa ^{2}:=8\pi .$} We obtain the Hilbert--Einstein action if
$f(R,T)=R.$ Such constructions are reviewed in \cite%
{nojiri,tsujikawa,harko,harko1}.

\subsubsection{The Finsler--Cartan gravity}

In the second class of theories, we consider instead of (\ref{qel}) a
nonlinear quadratic element,
\begin{eqnarray}
ds^{2} &=&F^{2}(x^{i},y^{j})  \label{nqel} \\
&\approx &-(cdt)^{2}+g_{\widehat{i}\widehat{j}}(x^{k})y^{\widehat{i}}y^{%
\widehat{j}}[1+\frac{1}{r}\frac{\rho _{\widehat{i}_{1}\widehat{i}_{2}...%
\widehat{i}_{2r}}(x^{k})y^{\widehat{i}_{1}}...y^{\widehat{i}_{2r}}}{\left(
g_{\widehat{i}\widehat{j}}(x^{k})y^{\widehat{i}}y^{\widehat{j}}\right) ^{r}}%
]+O(\rho ^{2}).  \notag
\end{eqnarray}%
for $y^{i}=dx^{i}/d\tau $ with a real parameter $\tau $ in $x^{i}(\tau ),$
where values $\rho _{\widehat{i}_{1}\widehat{i}_{2}...\widehat{i}_{2r}}(x)$
are parameterized by 3--d spacelike "hat" indices running values $\widehat{i}%
=1,2,3$ have to be computed using certain experimental/observational data
and/or theoretical models. To spacetime geometry and/or geometric mechanics,
locally anisotropic field theories with effective nonlinear metrics of type (%
\ref{nqel}), models of quantum gravity etc we can naturally associate \cite%
{laem,vmdrhl} certain local modified dispersion relations for propagation of
light. For a corresponding frequency $\omega $ and wave vector $k_{i},$ one
computes locally
\begin{equation}
\omega ^{2}=c^{2}[g_{\widehat{i}\widehat{j}}k^{\widehat{i}}k^{\widehat{j}%
}]^{2}(1-\frac{1}{r}\rho {_{\widehat{i}_{1}\widehat{i}_{2}...\widehat{i}%
_{2r}}y^{\widehat{i}_{1}}...y^{\widehat{i}_{2r}}}/{[g_{\widehat{i}\widehat{j}%
}y^{\widehat{i}}y^{\widehat{j}}]^{2r}}),  \label{disp}
\end{equation}%
when the local wave vectors $k_{i}\rightarrow p_{i}\sim y^{a}$ are related
to momentum type variables $p_{i}$ which are dual to "fiber" coordinates $%
y^{a}.$

Nonlinear metric elements (\ref{nqel}) are usually considered in Finsler
geometry when certain homogeneity conditions are imposed, $F(x^{i},\beta
y^{j})=\beta F(x^{i},y^{j}),$ for any $\beta >0).$ The value $F$ is
considered to be a fundamental (generating) Finsler function usually
satisfying the condition that the Hessian
\begin{equation}
\tilde{g}_{ij}(x^{i},y^{j})=\frac{1}{2}\frac{\partial ^2 F^{2}}{\partial
y^{i}\partial y^{j}}  \label{hess}
\end{equation}%
is not degenerate. For physical applications related to "small" deformations
of GR, we can consider that $g_{ij}=(-1,g_{\widehat{i}\widehat{j}}(x^{k}))$
in the limit $\rho \rightarrow 0$ correspond to a metric on a (pseudo)
Riemannian manifold with local coordinates $(x^{i})$ and signature of metric
of type $(-+++).$ In such cases, we elaborate (pseudo) Finsler models on
tangent bundles to Lorentz manifolds.

There are substantial differences between geometric and physical theories
constructed for (pseudo) Riemannian quadratic elements (\ref{qel}) and those
with nonlinear (Finsler type) ones (\ref{nqel}). In the first case, the data
$(g_{ij},\nabla )$ provide a complete geometric model for which gravity
theories are derived for corresponding Lagrange densities. \ We need more
assumptions in order to construct some self--consistent geometries from a
generating function $F(x,y).$ A metric compatible model of (pseudo)
Finsler--Cartan geometry completely determined by $F$ and $\tilde{g}_{ij}$,
up to necessary classes of frame/coordinate transform $e^{\alpha ^{\prime
}}=e_{\ \alpha }^{\alpha ^{\prime }}(x,y)e^{\alpha },$ can be constructed
from a triple $\left( F:\mathbf{N,g,D}\right) $ of fundamental geometric
objects:

\begin{enumerate}
\item The nonlinear connection (N--connection) structure:
\begin{equation}
\mathbf{N}:\ T\mathbf{TV}=h\mathbf{TV}\oplus v\mathbf{TV,}  \label{whit}
\end{equation}%
i.e. a nonholonomic (equivalently, non--integrable/ anholonomic)
distribution with horizontal (h) and vertical (v) splitting. This value can
be introduced in coefficient form, $\mathbf{N}=\{\mathbf{N}_{i^{\prime
}}^{a^{\prime }}=e_{\ a}^{a^{\prime }}e_{i^{\prime }}^{\ i}\tilde{N}%
_{i}^{a}\}$, where
\begin{equation*}
\tilde{N}_{j}^{a}:=\frac{\partial \tilde{G}^{a}(x,y)}{\partial y^{j}},%
\mbox{\ for \ }\tilde{G}^{k}=\frac{1}{4}\tilde{g}^{kj}\left( y^{i}\frac{%
\partial ^{2}L}{\partial y^{j}\partial x^{i}}-\frac{\partial L}{\partial
x^{j}}\right) .
\end{equation*}%
A N--adapted frame structure is defined naturally as $\mathbf{\tilde{e}}%
_{\nu }=(\mathbf{\tilde{e}}_{i},e_{a}),$ where
\begin{equation}
\mathbf{\tilde{e}}_{i}=\frac{\partial }{\partial x^{i}}-\tilde{N}_{i}^{a}(u)%
\frac{\partial }{\partial y^{a}}\mbox{ and
}e_{a}=\frac{\partial }{\partial y^{a}},  \label{dder}
\end{equation}%
and the dual frame (coframe) structure is $\mathbf{\tilde{e}}^{\mu }=(e^{i},%
\mathbf{\tilde{e}}^{a}),$ where
\begin{equation}
e^{i}=dx^{i}\mbox{ and }\mathbf{e}^{a}=dy^{a}+\tilde{N}_{i}^{a}(u)dx^{i};
\label{ddif}
\end{equation}%
The following nonholonomy relations are satisfied
\begin{equation}
\lbrack \mathbf{\tilde{e}}_{\alpha },\mathbf{\tilde{e}}_{\beta }]=\mathbf{%
\tilde{e}}_{\alpha }\mathbf{\tilde{e}}_{\beta }-\mathbf{\tilde{e}}_{\beta }%
\mathbf{\tilde{e}}_{\alpha }=\tilde{W}_{\alpha \beta }^{\gamma }\mathbf{%
\tilde{e}}_{\gamma }  \label{anhrel}
\end{equation}%
with anholonomy coefficients $\tilde{W}_{ia}^{b}=\partial _{a}\tilde{N}%
_{i}^{b}$ and $\tilde{W}_{ji}^{a}=\tilde{\Omega}_{ij}^{a}.$ \footnote{\label{fn4}
A N--connection can be canonically determined by $F$ following a
geometric/variational principle for an effective regular Lagrangian $L=F^{2}$
and action integral $S(\tau )=\int\limits_{0}^{1}L(x(\tau ),y(\tau ))d\tau ,%
\mbox{ for }y^{k}(\tau )=dx^{k}(\tau )/d\tau $. The Euler--Lagrange
equations $\frac{d}{d\tau }\frac{\partial L}{\partial y^{i}}-\frac{\partial L%
}{\partial x^{i}}=0$ are equivalent to the \textquotedblright nonlinear
geodesic\textquotedblright\ (equivalently, semi--spray) equations $\frac{%
d^{2}x^{k}}{d\tau ^{2}}+2\tilde{G}^{k}(x,y)=0$, where $\tilde{g}^{kj}$ is
inverse to $\ ^{v}{\tilde{g}}_{ij}\equiv {\tilde{g}}_{ij}$ (\ref{hess}).}

\item Using data $\left( {\tilde{g}}_{ij},\mathbf{\tilde{e}}_{\alpha
}\right) ,$ we can define a canonical (Sasaki type) metric structure
\begin{eqnarray}
\mathbf{\tilde{g}} &=&\tilde{g}_{ij}(x,y)\ e^{i}\otimes e^{j}+\tilde{g}%
_{ij}(x,y)\ \mathbf{\tilde{e}}^{i}\otimes \ \mathbf{\tilde{e}}^{j}
\label{slm} \\
&=&\ g_{ij}(x,y)\ e^{i}\otimes e^{j}+\ h_{ab}(x,y)\ \mathbf{e}^{a}\otimes
\mathbf{e}^{b},  \label{dm}
\end{eqnarray}%
which can be related to an "arbitrary" metric structure $\mathbf{g}=\{%
\mathbf{g}_{\alpha ^{\prime }\beta ^{\prime }}\}$ via frame transforms, $%
\mathbf{g}_{\alpha ^{\prime }\beta ^{\prime }}=e_{\ \alpha ^{\prime
}}^{\alpha }e_{\ \beta ^{\prime }}^{\beta }\mathbf{\tilde{g}}_{\alpha \beta
}.$

\item For any metric $\mathbf{g}=\tilde{\mathbf{g}}$ we can construct in
standard form the Levi--Civita connection $\nabla =\{\Gamma _{\beta \gamma
}^{\alpha }\},$ which does not preserve under parallelism the N--connection
splitting (\ref{whit}). In Finsler theories, one introduces distinguished
connections (d--connections) $\mathbf{D=\{\Gamma }_{\beta \gamma }^{\alpha }%
\mathbf{\}}$ which is adapted to the N--connection structure, i.e. preserves
the nonholonomic h-v--splitting. It is possible to construct
Einstein--Finsler type theories for d--connections with are compatible with
the metric structure, $\mathbf{Dg=0}$. For instance, this is the case of the
well known Cartan d--connection, which is metric compatible, but the Chern
and/or Berwald d--connections are not metric compatible which is less
related to standard models of physics, see discussions and critical remarks
in \cite{vcrit,vrflg}.
\end{enumerate}

Using N--adapted differential forms and the d--connection 1--form is $%
\mathbf{\Gamma }_{\ \beta }^{\alpha }=\mathbf{\Gamma }_{\ \beta \gamma
}^{\alpha }\mathbf{e}^{\gamma }$, we can define and compute the torsion and
curvature 2--forms,
\begin{eqnarray*}
\mathcal{T}^{\alpha }:= &&\mathbf{De}^{\alpha }=d\mathbf{e}^{\alpha }+%
\mathbf{\Gamma }_{\ \beta }^{\alpha }\wedge \mathbf{e}^{\beta }%
\mbox{\ and,
respectively, } \\
\mathcal{R}_{~\beta }^{\alpha }:= &&\mathbf{D\Gamma }_{\ \beta }^{\alpha }=d%
\mathbf{\Gamma }_{\ \beta }^{\alpha }-\mathbf{\Gamma }_{\ \beta }^{\gamma
}\wedge \mathbf{\Gamma }_{\ \gamma }^{\alpha }=\mathbf{R}_{\ \beta \gamma
\delta }^{\alpha }\mathbf{e}^{\gamma }\wedge \mathbf{e}^{\delta }.
\end{eqnarray*}%
For instance, the $h$--$v$--coefficients $\mathbf{T}_{\ \beta \gamma
}^{\alpha }=\{T_{\ jk}^{i},T_{\ ja}^{i},T_{\ ji}^{a},T_{\ bi}^{a},T_{\
bc}^{a}\}$ of $\mathcal{T}^{\alpha }$ are computed using formulas
\begin{eqnarray}
T_{\ jk}^{i} &=&L_{\ jk}^{i}-L_{\ kj}^{i},\ T_{\ ja}^{i}=-T_{\ aj}^{i}=C_{\
ja}^{i},\ T_{\ ji}^{a}=\Omega _{\ ji}^{a},\   \notag \\
T_{\ bi}^{a} &=&\frac{\partial N_{i}^{a}}{\partial y^{b}}-L_{\ bi}^{a},\
T_{\ bc}^{a}=C_{\ bc}^{a}-C_{\ cb}^{a};  \label{dtors}
\end{eqnarray}%
see, for instance, \cite{vrflg} \ for N--adapted coefficients of curvature, $%
\mathbf{R}_{\ \beta \gamma \delta }^{\alpha }$, and $$\Omega _{ij}^{a}=\frac{%
\partial N_{i}^{a}}{\partial x^{j}}-\frac{\partial N_{j}^{a}}{\partial x^{i}}%
+N_{i}^{b}\frac{\partial N_{j}^{a}}{\partial y^{b}}-N_{j}^{b}\frac{\partial
N_{i}^{a}}{\partial y^{b}}.$$

For our purposes (in order to decouple and integrate in very general forms
the gravitational field equations), it is convenient to work with the
so--called canonical d--connection $\mathbf{D}$ completely defined by a
metric $\mathbf{g=\tilde{g}}$ in metric compatible form, $\mathbf{Dg=}0,$
and with zero $h$- and $v$-torsions, $T_{\ jk}^{i}=0$ and $T_{\ bc}^{a}=0;$
in general, there are nonzero values $T_{\ ja}^{i},T_{\ ji}^{a}$ and $T_{\
bi}^{a},$ see (\ref{dtors})).\footnote{%
In our former works, we used the symbol $\widehat{\mathbf{D}}$ for the
canonical d--connection; here, we write the N--adapted coefficients of $%
\mathbf{D}$ are $\mathbf{\Gamma }_{\ \alpha \beta }^{\gamma }=\left(
L_{jk}^{i},L_{bk}^{a},C_{jc}^{i},C_{bc}^{a}\right) ,$
\begin{eqnarray*}
L_{jk}^{i} &=&\frac{1}{2}g^{ir}\left( \mathbf{e}_{k}g_{jr}+\mathbf{e}%
_{j}g_{kr}-\mathbf{e}_{r}g_{jk}\right),\ \widehat{C}_{bc}^{a}=\frac{1}{2}%
h^{ad}\left( e_{c}h_{bd}+e_{c}h_{cd}-e_{d}h_{bc}\right),  \notag \\
L_{bk}^{a} &=&e_{b}(N_{k}^{a})+\frac{1}{2}h^{ac}\left( e_{k}h_{bc}-h_{dc}\
e_{b}N_{k}^{d}-h_{db}\ e_{c}N_{k}^{d}\right),\ C_{jc}^{i} =\frac{1}{2}%
g^{ik}e_{c}g_{jk}.  \label{candcon}
\end{eqnarray*}%
} Here we note that there is a canonical distortion relation
\begin{equation}
\mathbf{D=}\nabla +\mathbf{Z}  \label{dcdc}
\end{equation}%
where both connections $\mathbf{D}$ and $\nabla $ and the distortion tensor $%
\mathbf{Z}$ (it is an algebraic combination of nontrivial torsion
coefficients, see explicit formulas in \cite{vrflg}) are uniquely defined by
the same metric structure $\mathbf{g.}$

We can construct a variant of Einstein--Finsler theory $\mathbf{D}$
following standard geometric rules as in general relativity but
reconsidering the constructions on tangent bundles/manifolds. The scalar
curvature is by definition
\begin{equation}
\ _{s}^{F}R:=\mathbf{g}^{\beta \gamma }\ \mathbf{\mathbf{R}}_{\beta \gamma
}=g^{ij}R_{ij}+h^{ab}R_{ab}=\ ^{h}R+\ ^{v}R.  \label{riccifs}
\end{equation}
This scalar curvature is similar to that for the Levi--Civita connection in the Einstein gravity. In both cases of a (pseudo) Riemannian geometry and/or a Finsler space, such a value is uniquely defined on the corresponding total tangent bundle by contracting the total metric tensor and the respective Riemannian tensor. Formulas are similar but with that difference that in the second case we consider  a Finsler like connection.

It should be noted that Finsler like variables can be introduced in standard
GR considering a generating function $F=\mathcal{F}(x,y)$ determining a 2+2
splitting. For such models, indices $i,j,...=1,2$ and $a,b,...=3,4$ which is
adapted to a fibred structure on a Lorentz manifold. We can use similar
geometric constructions with $4+4$ splitting when Finsler models are on
tangent bundles, and distinguish this via conventional $i,j,...=1,2,3,4$ and
$a,b,...=5,6,7,8.$

The gravitational field equations for $\mathbf{D}$ can be postulated in
standard geometric form and/or derived via N--adapted variational calculus,
\begin{eqnarray}
\mathbf{R}_{\ \beta \delta }-\frac{1}{2}\mathbf{g}_{\beta \delta }\ \
_{s}^{F}R &=&\mathbf{\Upsilon }_{\beta \delta },  \label{cdeinst} \\
L_{aj}^{c}=e_{a}(N_{j}^{c}),C_{jb}^{i}=0,\ \Omega _{\ ji}^{a} &=&0,
\label{lcconstr}
\end{eqnarray}%
for \ $\mathbf{\Upsilon }_{\beta \delta }\rightarrow T_{\beta \delta }$ if $%
\mathbf{D}\rightarrow \nabla .$ We have to consider the constraints (\ref%
{lcconstr}) in order to get zero torsion (\ref{dtors}) and distortion
tensors, $\mathbf{Z}=0,$ which constraints $\widehat{\mathbf{D}}=\nabla $ in
N--adapted frames, (\ref{dcdc}). It is convenient to work with equations of
type (\ref{cdeinst}) and (\ref{lcconstr}) if we wont to study in an unified
form both the Einstein gravity and Finsler generalized theories. Such
N--adapted Finsler like variables result into a very important property of
decoupling respective PDE which allows to construct solutions in very
general forms.

\subsubsection{The osculating approximation and $f(R,T,F)$ gravity}

Considering arbitrary frame/coordinate transforms on $\mathbf{V}$ and $T%
\mathbf{V}$, we mix the variables and do not "see" explicit dependencies on $%
F(x,y).$ Fixing a system of reference, we can introduce an osculating
(pseudo) Riemannian metric on the $h$--subspace%
\begin{equation}
\mathbf{g}_{ij}=\tilde{g}_{ij}(x,y(x)),  \label{oscul}
\end{equation}%
where $\tilde{g}_{ij}$ is defined by (\ref{hess}). In general, we can
construct exact solutions of (\ref{cdeinst}) for 8-d metrics (\ref{slm})
and/or (\ref{dm}). Nevertheless, the observable spacetime is four
dimensional and we can verify possible physical implications, directly, only
for the $h$--components. Any modifications via $F$ and $\mathbf{g}_{ij}$ (%
\ref{oscul}), and related nonholonomic deformations of $\ _{s}^{F}R$ (\ref%
{riccifs}) can be parameterized as $\ _{s}^{F}R\rightarrow f(\ ^{h}R,T,F),$
where $\ ^{h}R$ is computed for $\tilde{g}_{ij}.$

In explicit form, we can determine experimentally $F$ for theories with
modified dispersions (\ref{disp}) and, for instance, restricted local
Lorentz invariance, see reviews of results in \cite%
{mavromatos,sindoni,visser}. There are experimental restrictions for such
configurations \cite{laem}. Nevertheless, only local considerations are not
enough to conclude if a Finsler like theory is physically important, or not.
For instance, any data $\left( F:\mathbf{g,N,D}\right) $ can be redefined
equivalently via frame transforms into $\left( \ ^{0}F:\ ^{0}\mathbf{g,}\
^{0}\mathbf{N,}\ ^{0}\mathbf{D}\right) ,$ where $\ ^{0}F$ is a trivial
"Finsler" function resulting in quadratic element (\ref{qel}) but the data $%
\left(\ ^{0}\mathbf{g,}\ ^{0}\mathbf{N,}\ ^{0}\mathbf{D}\right) $ are
constructed as solutions of (\ref{cdeinst}) with possible, or not,
Levi--Civita constraints (\ref{lcconstr}). Experimentally, we shall obtain
quadratic dispersions in (\ref{nqel}) and (\ref{disp}) but the information
on locally anisotropic types (Finsler, or other types) is encoded into
N--adapted frames $\ ^{0}\mathbf{N}$ and generic off--diagonal terms of $\
^{0}\mathbf{g.}$ One could be observational effects for Finsler brane and
black hole/ellipsoid solutions with (non) commutative variables and
anisotropic modified dispersions \cite{vfbr,vmdrhl}.

The principles of generalized covariance can be extended from $\mathbf{V}$
to $T\mathbf{V}$. Various classes of exact solution of gravitational field
equations $T\mathbf{V}$ contains directly or indirectly contain physical
information on $F$. Such data can be encoded into N- and d--connections and
total metrics. Possible phenomenology and experimental/observational effects
can analyzed for metrics of type (\ref{oscul}) with certain nonholonomic
projections of theories on fundamental $h$--spacetime. Geometrically, we can
transform a diagonal configuration on $T\mathbf{V,}$ determined with respect
to N--adapted bases (\ref{dder}) and (\ref{ddif}), into a generic
off--diagonal $\mathbf{g}_{ij}[F,\tilde{g}_{ij}]$ when the functional
dependence can be stated for well--defined boundary/assymptotic conditions,
Cauchy problem etc. This class of theories is of type $f(R,T,F)$ $%
\rightarrow $ $\ ^{h}R[F,\tilde{g}_{ij}],$ where $F$ can be defined up to
certain classes of symmetries under coordinate/frame transforms and, in
special cases, reduced local symmetries.

Working with $\mathbf{g}_{ij}[F,\tilde{g}_{ij}]$, we use an effective $%
\mathcal{F}(x^{1},x^{2};y^{3},y^{4})$ generating conventional $2+2$
splitting which for corresponding N--adapted bases we can construct
off--diagonal solutions for 4--d gravity theories. In a particular case, we
can constrain the integral varieties of solutions in order to extract
Levi--Civita configurations. Such generating functions $\mathcal{F}$ can be
also determined up to certain classes of frame transforms. In all cases, we
can conventionally write $f(R,T,F)$ where $F $ emphasizes possible locally
anisotropic/ nonholonomic / generic off--diagonal contributions from certain
Finsler like models on $\mathbf{V}$, or $T\mathbf{V.}$

\subsection{Field equations for the $f(R,T,F)$ gravity}

Hereafter, we assume that small Greek indices split in the form $\alpha
,\beta ,...=(i,a),(j,b)...,$ where $i,j,...=1,2$ and $a,b=3,4$ for a
osculating (pseudo) Riemannian metric of type (\ref{oscul})
\begin{equation}
\mathbf{g}_{\alpha \beta }(u^{\gamma })=\mathbf{\tilde{g}}_{\alpha \beta
}(u^{\gamma },y^{\alpha }(u^{\mu }))  \label{osculef}
\end{equation}%
on a nonholonomic Lorentz manifold $\mathbf{V}$ with local coordinates $%
u^{\alpha }=(x^{i},y^{a})$ with functional dependence on sections $y^{\alpha
}(u^{\mu })$ of $T\mathbf{V.}$ We shall not state theoretical/experimental
constraints for $F(u^{\gamma },y^{\alpha })$ on open regions of $T\mathbf{V}$
but analyze general and physical important implications of such nontrivial
structures on effective (pseudo) Riemannian spacetime $\mathbf{V.}$ The
metric $\mathbf{g}_{\alpha \beta }(u^{\gamma })$ (\ref{osculef}) can be
parameterized in N--adapted from as (\ref{slm}) and/or (\ref{dm}), for 4--d
configurations with nonholonomic splitting into 2-d $h$-components and 2-d
v--components. It is possible to introduce Finsler like variables on $%
\mathbf{V}$ if we prescribe an effective nonholonomic distribution $%
F\rightarrow \mathcal{F}(x^{i},y^{a}).$

For simplicity, we assume that the Lagrangian density of matter$\ \mathcal{L}%
(u^{\gamma })$ depends only the metric tensor components $\mathbf{g}_{\alpha
\beta },$ when
\begin{equation}
\mathbf{T}_{\alpha \beta }\mathbf{=}-2(\sqrt{|\mathbf{g}|})^{-1}\delta (%
\sqrt{|\mathbf{g}|}\ \mathcal{L})/\delta \mathbf{g}^{\alpha \beta }=\mathbf{g%
}_{\alpha \beta }\mathcal{L}-2\partial \mathcal{L}/\partial \mathbf{g}%
^{\alpha \beta },  \label{emst1}
\end{equation}%
for $|\mathbf{g}|$ being the determinant of (\ref{osculef}). We denote by $%
\mathbf{T}:=\mathbf{T}_{\beta }^{\beta }.$ We note that N--adapted
variations are obtained with respect to N--elongated frames (\ref{dder}) and
(\ref{ddif}) assuming that we work on a spacetime with nonholonomic $2+2$
splitting.

Our modified gravity theory is modelled on $\mathbf{V}$ by a functional $f(R,%
\mathbf{T,}F)\simeq \ ^{h}R,$ for $R\simeq \ _{s}^{F}R$ as in (\ref{riccifs}%
). The action is considered in the forms
\begin{eqnarray}
S &=&\int \sqrt{|\mathbf{g}|}d^{2}x\delta ^{2}y\{(16\pi )^{-1}f(R,\mathbf{T,}%
F)+\mathcal{L}\}  \label{sec1} \\
&\simeq &\int \sqrt{|\mathbf{\tilde{g}}|}d^{2}x\delta ^{2}y\{\ _{s}^{F}R+%
\mathcal{L}\},  \label{sec2}
\end{eqnarray}%
where the term (\ref{sec2}) with $\ _{s}^{F}R$ is a part of theory on $T%
\mathbf{V,}$ which we do not state in explicit form. We shall formulated
certain physically important conditions for a theory for (\ref{sec1}) which
will be described by exact solutions of (\ref{cdeinst}) related to (\ref%
{sec2}). For The N--adapted variation of $S$ (computations are similar to
those for derivation of formulas (11) in  \cite{harko}, but performed for
the canonical d--connection $\mathbf{D}$ and metrics of type $\mathbf{g}$ (%
\ref{dm}), for $\partial f/\partial R=\partial _{R}f\neq 0),$ we obtain the
locally anisotropic gravitational field equations%
\begin{equation}
\mathbf{\mathbf{R}}_{\beta \gamma }-\frac{1}{2}\frac{f}{(\partial _{R}f)}%
\mathbf{g}_{\beta \gamma }=\frac{1}{\partial _{R}f}[8\pi \mathbf{T}_{\beta
\gamma }+(\mathbf{D}_{\beta }\mathbf{D}_{\gamma }-\mathbf{g}_{\beta \gamma }%
\mathbf{D}_{\alpha }\mathbf{D}^{\alpha })(\partial _{R}f)-(\partial _{T}f)(%
\mathbf{T}_{\beta \gamma }+\mathbf{\Theta }_{\beta \gamma })]  \label{mgfeq}
\end{equation}%
where
\begin{equation}
\mathbf{\Theta }_{\beta \gamma }:=\mathbf{g}^{\mu \nu }\delta (\mathbf{T}%
_{\mu \nu })/\delta \mathbf{g}^{\beta \gamma }\mbox{ \ and \ }\mathbf{\Theta
:=\Theta }_{\mu }^{\ \mu }.  \label{ems1}
\end{equation}%
It should be noted that the divergence of $\mathbf{T}_{\beta \gamma }$ is
not zero,
\begin{equation}
(8\pi (\partial _{T}f)^{-1}-1)\mathbf{D}_{\mu }\mathbf{T}^{\mu \nu }=(%
\mathbf{T}^{\mu \nu }+\mathbf{\Theta }^{\mu \nu })\mathbf{D}_{\mu }\ln
|\partial _{T}f|+\mathbf{D}_{\mu }\mathbf{\Theta }^{\mu \nu }.  \label{divem}
\end{equation}%
Such properties with $\mathbf{D}_{\mu }\mathbf{T}^{\mu \nu }\neq 0$ are
known in Finsler gravity theories and GR in nonholonomic variables, see \cite%
{vrflg}, when $\mathbf{D}_{\mu }$ is of type (\ref{dcdc}) with all
components determined by $\mathbf{g},$ and if $\mathbf{D}_{\mu }\rightarrow
\nabla _{\mu },\nabla _{\mu }\mathbf{T}^{\mu \nu }=0.$

Finally, we note that the equations \ (\ref{mgfeq}) are similar to (\ref%
{cdeinst}) with effective sources $\mathbf{\Upsilon }_{\beta \delta }$
depending on the physical nature of matter fields determined by $\mathbf{%
\Theta }_{\beta \gamma }.$

\section{Particular Cases and Effective Gravity Models}

\label{spc}In this section, we consider several classes of modified gravity
theories with explicit parametrization for sources and functional $f.$ We
shall analyze the possibility to reconstruct gravity with scalar field and
off--diagonal interactions. We will study the effective locally anisotropic
motion and Newtonian limits. We also will provide the Raychaudhuri equation
on the tangent Lorentz bundle $TV$.

\subsection{Effective Finsler--like and Einstein configurations}

\subsubsection{Assumptions on stress--energy tensors}

The calculation of $\mathbf{\Theta }_{\beta \gamma }$ is possible if the
matter Lagrangian is postulated. Using formulas (\ref{emst1}) and (\ref{ems1}%
), we find%
\begin{equation}
\mathbf{\Theta }_{\beta \gamma }=\mathbf{g}_{\beta \gamma }\mathcal{L}-2%
\mathbf{g}^{\alpha \tau }\partial ^{2}\mathcal{L}/\partial \mathbf{g}%
^{\alpha \tau }\partial \mathbf{g}^{\beta \gamma }-2\mathbf{T}_{\beta \gamma
}.  \label{ems2}
\end{equation}%
There are three such important models of matter fields subjected to
nonholonomic constraints:

\begin{enumerate}
\item For perfect fluids, we assume that with respect to N--adapted frames
the four--velocity field satisfy the conditions $\mathbf{v}_{\alpha }\mathbf{%
v}^{\alpha }=1$ and $\mathbf{v}^{\alpha }\mathbf{D}_{\mu }\mathbf{v}_{\alpha
}=0.$ There is not a unique definition but we shall take the Lagrangian
density $L=-p.$ For conventional energy density $\rho $ and pressure $p,$ we
can parameterize
\begin{equation}
\mathbf{T}_{\beta \gamma }=(\rho +p)\mathbf{v}_{\beta }\mathbf{v}_{\gamma }-p%
\mathbf{g}_{\beta \gamma },  \label{pemt}
\end{equation}
when (\ref{ems2}) is computed $\mathbf{\Theta }_{\beta \gamma }=-2\mathbf{T}%
_{\beta \gamma }-p\mathbf{g}_{\beta \gamma }.$

\item We can consider a scalar field $\varphi (x,y)$ with zero mass, with
Lagrange density $\ ^{\varphi }\mathcal{L}=\mathbf{g}^{\alpha \tau }(\mathbf{%
D}_{\alpha }\varphi )(\mathbf{D}_{\tau }\varphi )$, when $\mathbf{\Theta }%
_{\beta \gamma }=-\ ^{\varphi }\mathbf{T}_{\beta \gamma }+(1/2)\ ^{\varphi }%
\mathbf{Tg}_{\beta \gamma }.$

\item A different relation, $\mathbf{\Theta }_{\beta \gamma }=-\ ^{F}\mathbf{%
T}_{\beta \gamma },$ is computed for the electromagnetic (antisymmetric)
tensor field $\mathbf{F}_{\alpha \beta }$ with the Lagrangian density, $\
^{F}\mathcal{L}=-(16\pi )^{-1}\mathbf{g}^{\alpha \tau }\mathbf{g}^{\beta
\gamma }\mathbf{F}_{\alpha \beta }\mathbf{F}_{\tau \gamma }.$
\end{enumerate}

The assumptions above on stress--energy fields are based on the principle of
general covariance in GR when the formulas for gravity--matter field
interactions are the same with respect to arbitrary frames of reference.
Possible contributions from modified $f$--terms, extra $v$--dimensions and
local anisotropies are encoded in N--adapted frames of reference and
d--connection $\mathbf{D.}$

\subsubsection{Models with $f(R,T,F)=\ ^{h}R+2f(T)$}

The scalar curvature $\ ^{h}R$ is the first term in (\ref{riccifs}), in 8-d,
taken for $\mathbf{g}_{\alpha \beta }(u^{\gamma })$ (\ref{osculef}), reduced
to 4--d, and $f(T)$ is an arbitrary function taken for the traces of the
stress--energy tensor of matter. We identify $\ ^{h}R$ with the scalar
curvature $\ \ _{s}^{F}R$ of $\mathbf{D}$ adapted to a 2+2 N--connection
splitting via a prescribed generating function $\mathcal{F}(u)$. The
gravitational field equations \ (\ref{mgfeq}) transform into a variant for
Finsler gravity, see (\ref{cdeinst}), with source
\begin{equation*}
\mathbf{\Upsilon }_{\beta \delta }\left[ f(T),\mathbf{T}_{\alpha \beta },%
\mathbf{\Theta }_{\alpha \beta }\right] =f(T)\mathbf{g}_{\beta \delta }+%
\left[ 8\pi -2\partial _{T}f(T)\right] \mathbf{T}_{\beta \delta }-2\partial
_{T}f(T)\mathbf{\Theta }_{\beta \delta }.
\end{equation*}%
Such a 4--d effective gravity model transforms into a theory for $\nabla $
if the conditions (\ref{lcconstr}) for zero torsion are imposed. In both
cases of connections $\mathbf{D}$ and/or $\nabla $, the functional
dependence $\mathbf{\Upsilon }_{\beta \delta }\left[ f(T),\mathbf{T}_{\alpha
\beta},\mathbf{\Theta }_{\alpha \beta }\right] $ does not allow to obtain
standard Einstein manifolds even we neglect terms with $\mathbf{T}_{\beta
\delta }$ and $\mathbf{\Theta }_{\beta \delta }.$ The term $f(T)\mathbf{g}%
_{\beta \delta }$ mimic a locally anisotropic polarization resulting from $%
T(u)$ of a gravitational constant $\lambda ,$ when, for instance, $%
f(T)=\lambda T(u).$ Such generic off--diagonal solutions were studied in a
series of our works, see review \cite{vrflg}.

For trivial N--connection structure, the constructions presented in this
subsection reduce to those analyzed in section III.A. of \cite{harko1}.
Nevertheless, there are known locally anisotropic cosmological
configurations with nonzero N--connection coefficients, i.e. off--diagonal
generalizations of FRWL universes studied in \cite{vofdc}. In all such
cases, we can consider perfect fluid or dust universe approximations and
construct cosmologies with effective cosmological constant and locally
anisotropic polarizations. The generic off--diagonal solutions can be
constructed with generalized group symmetries with may contain information
on symmetries for the Finsler generating function $F$ in some 8--d models.

\subsubsection{Modified theories with $f(R,T,F)=\ ^{1}f(\ _{s}^{F}R)+\
^{2}f(T)$}

This is an example when the effective 4--d gravity is geometrically more
"sensitive" to Finsler contributions. For simplicity, assuming a matter
content for a perfect fluid, the field equations \ (\ref{mgfeq}) \ for $f=\
^{1}f(\ _{s}^{F}R)+\ ^{2}f(T)$ \ are reformulated in the form (\ref{cdeinst}%
) with modified gravitational constant and effective source
\begin{equation*}
\mathbf{\Upsilon }_{\beta \delta }=8\pi \ ^{ef}G\ \mathbf{T}_{\beta \gamma
}+\ ^{ef}\mathbf{T}_{\beta \gamma },
\end{equation*}%
where the effective values are computed
\begin{eqnarray*}
\ ^{ef}G &=&\left[ 1+(8\pi )^{-1}\partial _{T}(\ ^{2}f)\right] /\partial
_{R}(\ ^{1}f), \\
2\partial _{R}(\ ^{1}f)\ ^{ef}\mathbf{T}_{\beta \gamma } &=&\left[ \left(
1-2\partial _{R}\right) (\ ^{1}f)+(1+2\partial _{T})(\ ^{2}f)\right] \mathbf{%
g}_{\beta \gamma }+2(\mathbf{D}_{\beta }\mathbf{D}_{\gamma }-\mathbf{g}%
_{\beta \gamma }\mathbf{D}^{\alpha }\mathbf{D}_{\alpha })\partial _{R}(\
^{1}f).
\end{eqnarray*}%
We can compute variations of $\ ^{ef}G$ by a fundamental Finsler function $F$
in tangent Lorentz bundle even we follow an osculating approximation to
4--d. If we impose the conditions (\ref{lcconstr}), we get a Levi--Civita
configuration but with effective (matter and time) dependent coupling. In
general, locally anisotropic Finsler like contributions are "inverse" ones
comparing to matter modifications.

In the scenario with $\ ^{ef}G$ and $\ ^{ef}\mathbf{T}_{\beta \gamma },$ the
cosmic acceleration may result in three possible forms: with locally
anisotropic Finsler like contributions, depending on matter content of the
universe (for the matter and geometry coupling etc and modification of the
Hilbert--Einstein terms in Lagrange density) and via an effective source
term in the right part of effective Einstein equations. It should be noted
here that such nonholonomic deformations of theories are with generic
N--connection structure.

Of course, we can consider another types of parametrization of $f(R,T,F)$
resulting in different classes of effective sources and off--diagonal
deformations. If a generalized principle of relativity is considered for
such classes of theories, we can model such theories as certain branches of
nonholonomic manifolds/bundles geometries. We can use this for defining such
transforms when the effective gravitational equations decouple and can be
integrated in general forms.

\subsection{Equivalence of models with $f(R,T,F)$ to effective GR or
Einstein--Finsler gravity}

Frame and conformal transforms change the geometric and matter field
components of theories when a theory of type (\ref{sec1}) can be modelled as
(\ref{sec2}) and/or inversely. The gravitational field equations are also
modified (\ref{mgfeq}) both via functionals $f(...)$ and $F(...).$ For
simplicity, we consider the first action for $f=\ _{s}^{F}R+\ ^{2}f(T),$
where $\ _{s}^{F}R$ is of type (\ref{riccifs}), $q_{A}$ label matter fields
with energy--momentum tensor%
\begin{equation}
\mathbf{T}_{\alpha \beta }=-2(\sqrt{|g|})^{-1}\delta \int d^{4}x\sqrt{|%
\mathbf{g}|}\ \mathcal{L}(\mathbf{g}_{\alpha \beta },q_{A})/\delta \mathbf{g}%
_{\alpha \beta }.  \label{emst}
\end{equation}%
computed on the 4--d (h--part) base spacetime. The v--components of the
Ricci tensor \ in 8--d are stated equal to an effective polarized
cosmological constant $\ ^{v}R=\Lambda (x)$, we considered such models in
\cite{vfbr}. The field equations are of type (\ref{cdeinst}), which in 4--d
result in
\begin{equation*}
\mathbf{\mathbf{R}}_{\beta \gamma }-\frac{1}{2}\mathbf{g}_{\beta \gamma }\
_{s}^{F}R=\left[ 8\pi +\partial _{T}(\ ^{2}f)\right] \mathbf{T}_{\beta
\gamma }+\ ^{ef}\mathbf{T}_{\beta \gamma },
\end{equation*}%
with effective gravitational constant,
\begin{equation}
\ ^{ef}G=\left[ 8\pi +\partial _{T}(\ ^{2}f)\right] ,  \label{effgc1}
\end{equation}
and effective energy--momentum tensor, $\ ^{ef}\mathbf{T}_{\beta \gamma }=[\
^{2}f+\partial _{T}(^{2}f)]\mathbf{g}_{\beta \gamma }.$ Putting all terms
together, we get
\begin{equation}
\mathbf{\mathbf{R}}_{\beta \gamma }-\frac{1}{2}\left[ \ ^{h}R+\Lambda (x)+\
2\left( \ ^{2}f+2\partial _{T}(\ ^{2}f)\right) \right] \mathbf{g}_{\beta
\gamma }=\left[ 8\pi +\partial _{T}(\ ^{2}f)\right] \mathbf{T}_{\beta \gamma
}  \label{eeff1}
\end{equation}%
for the canonical d--connection $\mathbf{D},$ when the term $\ ^{2}f(T)$
modifies both the gravitational constant and may compensate a cosmological
constant $\Lambda =\Lambda _{0},$ or polarizations to $\Lambda (x)$ and
possible contributions by Finsler modifications in $\ ^{h}R.$

The theory described by (\ref{eeff1}) becomes an effective Einstein like
theory if we impose the Levi--Civita conditions (\ref{lcconstr}), $\mathbf{%
D\rightarrow \nabla ,}$ and fix such a parametrizations where $\
^{h}R+\Lambda (x)+\ 2\left( \ ^{2}f+2\partial _{T}(\ ^{2}f)\right) =\
^{\nabla }R,$ where $\ ^{\nabla }R$ is the scalar curvature of $\nabla .$
Nevertheless, variations of effective gravitational constant (\ref{effgc1})
are still possible. The gravitational field equations can be integrated in
very general off--diagonal forms following methods elaborated in \cite%
{veyms,vfbr}.

\subsection{Extracting scalar fields from modified/Finsler gravity}

\subsubsection{Conformal transforms \& effective scalar fields}

We show how the Finsler generating function $F(x,y(x)):=\chi (x)$ can be
used to mimic scalar interactions in modified gravity with conformal
transforms $\mathbf{g}_{\alpha \beta }\rightarrow \ ^{\chi }\mathbf{g}%
_{\alpha \beta }=e^{\chi (x)}\mathbf{g}_{\alpha \beta }.$ Let us consider \
a functional $f(\ _{s}^{F}R,\chi )$ being an algebraic function on $\
_{s}^{F}R$ and $\chi .$ We introduce the action
\begin{equation}
S=\int d^{4}x\sqrt{|\mathbf{g}|}\{\frac{1}{2\kappa ^{2}}f(\ _{s}^{F}R,\chi )+%
\mathcal{L}(e^{\chi (x)}\mathbf{g}_{\alpha \beta },q_{A})\}.  \label{act4}
\end{equation}%
Varying on $\chi ,$ we get the relation, $\partial _{\chi }f(\
_{s}^{F}R,\chi )=-\kappa ^{2}\ ^{\chi }\mathbf{T}$, where the trace of
energy--momentum tensor for the effective scalar field is computed $\ ^{\chi
}\mathbf{T}:= \ ^{\chi }\mathbf{g}^{\alpha \beta }\ ^{\chi }\mathbf{T}%
_{\alpha \beta }$, see formula (\ref{emst}) for re--scaled metric $e^{\chi
(x)}\mathbf{g}_{\alpha \beta }$, which results in $\ ^{\chi }\mathbf{T}%
_{\alpha \beta }.$

There are parametrization where we can invert and find $\chi =\chi (\
_{s}^{F}R,\ ^{\chi }\mathbf{T})$ and then re--define $\widehat{f}(\
_{s}^{F}R,\ ^{\chi }\mathbf{T})\equiv f(\ _{s}^{F}R,\chi (\ _{s}^{F}R,\
^{\chi }\mathbf{T})).$ Under conformal N--adapted transform $\mathbf{g}%
_{\alpha \beta }\rightarrow e^{-\chi (x)}\mathbf{g}_{\alpha \beta },$ we
induce from a modified Finsler action (\ref{act4}) an action for the
canonical d--connection $\mathbf{D}$ modified gravity with effective scalar
fields $\chi ,$%
\begin{equation}
S=\int d^{4}xe^{-2\chi (x)}\sqrt{|\mathbf{g}|}\{\frac{1}{2\kappa ^{2}}f(\
_{s}^{\chi }R\ ,\chi )+\mathcal{L}(\mathbf{g}_{\alpha \beta },q_{A})\},
\label{act4a}
\end{equation}%
\begin{equation*}
\mbox{ where \ } \ _{s}^{\chi }R=e^{\chi (x)}\left[ \ _{s}^{F}R+3\mathbf{g}%
^{\alpha \beta }\left( (\mathbf{D}_{\alpha }\chi )(\mathbf{D}_{\beta }\chi )-%
\frac{1}{2}(\mathbf{e}_{\alpha }\chi )(\mathbf{e}_{\beta }\chi )\right) %
\right] ,
\end{equation*}%
when the N--elongated operators $\mathbf{e}_{\alpha }$ are certain frame
transforms of (\ref{dder}) adapted to a nonholonomic 2+2 splitting. Such
bases are also physical frames if the matter fields $q_{A}$ couple minimally
with $\mathbf{g}_{\alpha \beta }$ but not with $\chi $ and the action is
re--written in the form%
\begin{equation}
S=\int d^{4}xe^{-2\chi (x)}\sqrt{|\mathbf{g}|}\{\frac{1}{2\kappa ^{2}}%
\widehat{f}(\ _{s}^{\chi }R\ ,\mathbf{T})+\mathcal{L}(\mathbf{g}_{\alpha
\beta },q_{A})\}  \label{act4b}
\end{equation}%
where $\mathbf{T=}\ ^{\chi }\mathbf{T+}\ ^{q}\mathbf{T}$ is computed for
sets of fields $(\chi ,q_{A}).$

\subsubsection{FLRW--geometries induced by Finsler modifications}

We briefly analyze possible cosmological implications of the models (\ref%
{act4a}) and/or (\ref{act4b}) when $\mathcal{L}=0$ and $f(\ _{s}^{\chi }R\
,\ ^{\chi }\mathbf{T})=\ _{s}^{F}R+f(\ ^{\chi }\mathbf{T})$ with a
corresponding re--definition of $F(x,y(x)):=\chi (x)$ to get an effective
\footnote{%
we can introduce, for instance, a term $-4\mathcal{V}(\chi )$ for nonlinear
interactions, as in various cosmological models; for simplicity, we restrict
our considerations to models with $\mathcal{V}(\chi )=0$}  $\ ^{\chi }%
\mathbf{T}=-\omega (\chi )\mathbf{g}^{\alpha \beta }(\mathbf{D}_{\alpha
}\chi )(\mathbf{D}_{\beta }\chi )-4V(\chi )$.  Such a Finsler modified
gravity theory,%
\begin{equation*}
S=\int d^{4}x\sqrt{|\mathbf{g}|}\{\frac{1}{2\kappa ^{2}}\
_{s}^{F}R+f[-\omega (\chi )\mathbf{g}^{\alpha \beta }(\mathbf{D}_{\alpha
}\chi )(\mathbf{D}_{\beta }\chi )]-\frac{1}{2}\omega (\chi )\mathbf{g}%
^{\alpha \beta }(\mathbf{D}_{\alpha }\chi )(\mathbf{D}_{\beta }\chi )\}
\end{equation*}%
contains (in our case, induced) $k$--essence cosmology models studied in
\cite{chiba,arm,mats,bamba1}. This follows from the Friedman equations
\begin{eqnarray}
\frac{3}{2\kappa ^{2}}H^{2} &=&-f[\Phi ]+\frac{1}{2}\Phi -\left( 2\partial
_{\Phi }f[\Phi ]-\Phi \right) \Phi ,  \notag \\
\frac{1}{\kappa ^{2}}(3H^{2}+2\dot{H}) &=&f[\Phi ]-\frac{1}{2}\Phi ,
\label{fried}
\end{eqnarray}%
for signature $(+,+,+,-),$ where $\Phi =\omega (\chi )\dot{\chi}^{2}$ is
computed using the derivative on time like variable $t,$ $\dot{\chi}%
=\partial \chi /\partial t.$

We can construct a simple solution for the model
\begin{equation*}
A(\chi ) =f[-\omega (\chi )\mathbf{g}^{\alpha \beta }(\mathbf{D}_{\alpha
}\chi )(\mathbf{D}_{\beta }\chi )]-\frac{1}{2}\omega (\chi )\mathbf{g}%
^{\alpha \beta }(\mathbf{D}_{\alpha }\chi )(\mathbf{D}_{\beta }\chi )
=A_{0}\exp [-2\ln (\chi /\chi _{0})\mathbf{g}^{\alpha \beta }(\mathbf{D}%
_{\alpha }\chi )(\mathbf{D}_{\beta }\chi )],
\end{equation*}%
with some constants $A_{0}$ and $\chi _{0}.$ The solution of (\ref{fried})
is very similar to that presented in \cite{harko1}, $H=H_{0}/t,\chi =t,$
when $3H_{0}^{2}-2H_{0}+(\kappa \chi _{0})^{2}A_{0}=0.$ In general, $k$%
--essence models, there are not de Sitter type solutions if (in our
denotations) $\chi =const$ and $A(0)>0.$

It is also interesting to note that physical implications of Finsler
modifications in gravity via functionals $f(...,F)$ may be very different
than those in "standard" $k$--essence cosmology. For instance, the
constraint $F(t,y(t))=\chi (t)$ $=const$ defines a nonholonomic distribution
in spacetime which may be "fiber like" with formal extra--dimensional
velocity coordinates $y^{i}(t),$ for a $2+2$ distribution as we explained in
footnote \ref{foot2}. The evolution in time depends on such parametrization.
Pseudo--Riemannian configurations can be obtained if $\mathbf{D\rightarrow
\nabla }$ (\ref{lcconstr}).

\subsection{Locally anisotropic motion and the Newtonian limit}

In $f(R,T,F)$ models, the divergence of energy--momentum tensor of matter (%
\ref{divem}) is not zero. For theories which can be modelled as an effective
Einstein gravity, this is a consequence of distortion $\mathbf{D=}\nabla +%
\mathbf{Z}$ (\ref{dcdc}). Nontrivial distributions $F$ result in
transferring solutions of $f(R,T)$ theories into Finsler like models and
off--diagonal Einstein configurations. The coupling between metric, geometry
and constraints on nonlinear dynamics induces supplementary accelerations
acting on test particles. The goal of this section, is to study the equation
of motion of test particles in dependence of both $f$-- and $F$--functionals
and distortions $\mathbf{Z.}$ We shall derive the equations of motion,
compute the Newtonian limits and investigate constraints on such theories
which can be derived from the observational data.

\subsubsection{Modified equations for anisotropic motion of test particles}

We compute the divergence (\ref{divem}) for the case of perfect fluid model
energy--momentum tensor (\ref{pemt}). Introducing the projector operator $\
^{\perp }\mathbf{g}_{\mu \lambda }=$ $\mathbf{g}_{\mu \lambda }-\mathbf{v}%
_{\mu }\mathbf{v}_{\lambda },$ for which $\ ^{\perp }\mathbf{g}_{\mu \lambda
}\mathbf{v}^{\mu }=0$ and $\ ^{\perp }\mathbf{g}_{\mu \lambda }\mathbf{T}%
^{\lambda \nu }=-\ \ ^{\perp }\mathbf{g}_{\lambda }^{\nu }p,$ and following
a calculus with $\mathbf{D}_{\alpha }$ and decompositions with respect to
N--adapted frames $\mathbf{e}_{\nu }$ (\ref{dder}) and $\mathbf{e}^{\nu }$ (%
\ref{ddif}) (see similar details for $\nabla $ in Section V of \cite{harko1}%
), we obtain that the equations of motion of a particle in background $(%
\mathbf{g,D=\{\mathbf{\Gamma }_{\ \nu \lambda }^{\mu }\}})$ can be expressed
\begin{equation}
\frac{d^{2}u^{\mu }}{ds^{2}}+\mathbf{\Gamma }_{\ \nu \lambda }^{\mu }\mathbf{%
v}^{\nu }\mathbf{v}^{\lambda }=(\mathbf{g}^{\mu \nu }-\mathbf{v}^{\nu }%
\mathbf{v}^{\mu })\mathbf{e}_{\nu }q.  \label{modifgeod}
\end{equation}%
The term $\mathbf{e}_{\nu }q$ can be found from divergence (\ref{divem}),
\begin{equation}
\mathbf{e}_{\nu }q=8\pi (\mathbf{e}_{\nu }p)(\rho +p)^{-1}[8\pi +\partial
_{T}f(\ _{s}^{F}R,\mathbf{T,}F)]^{-1}.  \label{aux5e}
\end{equation}

The equation (\ref{aux5e}) can be integrated using approximative methods and
additional assumptions on the matter fluid model. \ For \ instance, we can
chose a linear barotropic equation of state, $p=w\rho ,$ with a constant $%
w\ll 1,$ when $\rho +p\approx \rho $ and $\mathbf{T=}\rho -3p.$ The value $%
\partial _{T}f$ depends only on $\rho $ and $\ F$ and the deviations from
geodesic motion are determined by a term $\partial _{T}f=\ ^{1}\xi (\rho )+\
^{2}\xi \,(\rho ,F).$ Such contributions split into two terms: $\ ^{1}\xi $
derived from modifications of type $f(R,T)$ and $\ ^{2}\xi $ derived from an
anisotropic $F.$ Fixing a value $\rho =\rho _{0},$ we can expend%
\begin{equation*}
\partial _{T}f = \ ^{1}\xi (\rho _{0})+\ ^{2}\xi \,(\rho _{0},F)+(\rho -\rho
_{0})\left[ \frac{\partial \ ^{1}\xi }{\partial \rho }_{\mid \rho _{0}}d\rho
+\frac{\partial \ ^{2}\xi }{\partial \rho }_{\mid \rho _{0}}d\rho \right] =
8\pi \left[ \ ^{1}a+\ ^{2}a(F)+(\ ^{1}b+\ ^{2}b(F))(\rho -\rho _{0})\right] ,
\end{equation*}%
where $\ ^{1}a=\ ^{1}\xi /8\pi ,\ ^{1}b=\frac{\partial \ ^{1}\xi }{\partial
\rho }_{\mid \rho _{0}}d\rho $ and the anisotropic (Finsler generating
depending generating function coefficients) are $\ ^{2}a(F)=\ ^{2}\xi
\,/8\pi ,\ ^{2}b=\frac{\partial \ ^{2}\xi }{\partial \rho }_{\mid \rho
_{0}}d\rho .$ We can write (\ref{aux5e}) in the form
\begin{equation*}
\lbrack\ ^{1}a+\ ^{2}a(F)-(\ ^{1}b+\ ^{2}b(F))\rho _{0}]\mathbf{e}_{\nu}q=w%
\mathbf{e}_{\nu }\ln \{\rho /[\ ^{1}a+\ ^{2}a(F)+(\ ^{1}b+\ ^{2}b(F))(\rho
-\rho _{0})]\}
\end{equation*}
and get the approximate solution
\begin{equation}
q=\ln \left\{ \left[ \frac{C\rho }{\ ^{1}a+\ ^{2}a(F)+(\ ^{1}b+\
^{2}b(F))(\rho -\rho _{0})}\right] ^{w/[\ ^{1}a+\ ^{2}a(F)-(\ ^{1}b+\
^{2}b(F))\rho _{0}]}\right\} ,  \label{sol5e}
\end{equation}
where $C$ is an integration constant.

The solution (\ref{sol5e}) depends parametrically on generating Finsler
function via $\ ^{2}a(F)$ and $\ ^{2}b(F)$ and a logarithmic anisotropic
variations on \ $\rho .$ Such solutions can not be expressed in exact form
and there are necessary certain approximate series decompositions.

We extended the Raychaudhuri equations by using $\mathbf{D}$ connections in
the framework of a tangent Lorentz bundle $T\mathbf{V}$. We consider a
non-linear congruence of geodesics (\ref{modifgeod}) and we use an analogous
method with \cite{stavray} for deriving of Raychaudhuri equations, for a
velocity field $\mathbf{v}^{\mu }$ on $T\mathbf{V}$ the commutation
relations of $\mathbf{v}^{\mu }$ gives us
\begin{equation*}
\mathbf{D}_{\alpha }\mathbf{D}_{\beta }\mathbf{v}^{\gamma }-\mathbf{D}%
_{\beta }\mathbf{D}_{\alpha }\mathbf{v}^{\gamma }=\mathbf{R}_{\ \epsilon
\alpha \beta }^{\gamma }\mathbf{v}^{\epsilon }-\mathbf{T}_{\alpha \beta
}^{\delta }\mathbf{D}_{\delta }\mathbf{v}^{\gamma }-\mathbf{T}_{\alpha \beta
}^{\epsilon }\mathbf{D}_{\epsilon}\mathbf{v}^\gamma .
\end{equation*}%
Imposing the relations $\mathbf{g} ^{\perp} _{\mu \nu }\mathbf{v}^{\mu }=0$
and $\mathbf{v}^{\alpha }\mathbf{D}_{\beta }\mathbf{v}_{\alpha }=0,$ the
previous equation is written

\begin{equation*}
\mathbf{v}^{\beta }\mathbf{D}_{\alpha }\mathbf{D}_{\beta }\mathbf{v}^{\gamma
}=-\mathbf{v}^{\beta }\mathbf{v}_{\alpha }\mathbf{D}_{\beta }\mathbf{v}%
^{\gamma }+\mathbf{R}_{\ \epsilon \alpha \delta }^{\gamma }\mathbf{v}%
^{\epsilon }\mathbf{v}^{\delta }-\mathbf{T}_{\beta \alpha }^{\delta }\mathbf{%
D}_{\delta }\mathbf{v}^{\beta }\mathbf{v}^{\gamma }-\mathbf{T}_{\alpha \beta
}^{\epsilon }\mathbf{v}^{\beta }\mathbf{D}_{\epsilon }\mathbf{v}^{\gamma }.
\end{equation*}%
We decompose the term $\mathbf{D}_{\alpha }\mathbf{v}^{\beta }$ with respect
to the kinematics terms of expansion, shear and vorticity $\theta ,\sigma
,\omega $ as

\begin{equation*}
\mathbf{D}_{\alpha }\mathbf{v}^{\beta }=\frac{1}{7}\theta \mathbf{h}_{\alpha
}^{\beta }+\sigma _{\alpha }^{\beta }+\omega _{\alpha }^{\beta }
\end{equation*}%
where, correspondingly, $\theta =\mathbf{D}_{\alpha }\mathbf{v}^{\beta }%
\mathbf{h}_{\beta }^{\alpha },\ \sigma _{\alpha \beta }=\mathbf{D}_{\alpha }%
\mathbf{v}_{\beta }+\mathbf{D}_{\beta }\mathbf{v}_{\alpha }-\frac{1}{7}%
\theta \mathbf{h}_{\alpha \beta },\omega _{\alpha \beta }=\mathbf{D}_{\alpha
}\mathbf{v}_{\beta }-\mathbf{D}_{\beta }\mathbf{v}_{\alpha },\ \mathbf{h}%
_{\beta }^{\alpha }=\mathbf{g}^{\alpha \gamma }\mathbf{h}_{\beta \gamma }$.
We obtain
\begin{eqnarray*}
\mathbf{v}^{\alpha }\mathbf{D}_{\alpha }\theta &=&\mathbf{R}_{\alpha \beta }%
\mathbf{v}^{\alpha }\mathbf{v}^{\beta }-\mathbf{T}_{\alpha \gamma }^{\delta }%
\mathbf{v}^{\alpha }\mathbf{D}_{\delta }\mathbf{v}^{\gamma }-\mathbf{D}%
_{\alpha }\mathbf{v}^{\beta }\ \mathbf{D}_{\beta }\mathbf{v}^{\alpha } \\
&=&\mathbf{R}_{\alpha \beta }\mathbf{v}^{\alpha }\mathbf{v}^{\beta }-\mathbf{%
T}_{\alpha \beta }^{\delta }\left( \frac{1}{3}\theta \mathbf{h}_{\delta
}^{\beta }+\sigma _{\delta }^{\beta }+\omega _{\delta }^{\beta }\right)
\mathbf{v}^{\alpha }-\frac{1}{3}\theta -\sigma _{\beta }^{\alpha }\sigma
_{\alpha }^{\beta }-\omega _{\beta }^{\alpha }\omega _{\alpha }^{\beta }.
\end{eqnarray*}%
These equations are the (modified) Raychaudhuri equations on a Lorentz
tangent bundle. For vanishing nonholonomically induced torsion structures,
they transform into the well known Raychaudhuri equations for the
Levi--Civita connection but on the tangent bundle to a pseudo--Riemannian
manifold.

Using two test particles, we can study spacetime structure in Finsler like gravity theories using (modified) Raychaudhuri equations. There are possible also observable locally anisotropic effects on a single test particle which do not follow standard geodesic equations as in general relativity but certain nonlinear geodesic ones (see footnote \ref{fn4}) and formulas (\ref{modifgeod}) for modified equations of motion of a test particle in a Finsler like background.

\subsubsection{Corrections to the Newton law and perihelion effects}

We use pressureless dust and associated variational principle to study
modifications and anisotropies in the Newtonian limit of theories. The
equations for nonlinear geodesics (\ref{modifgeod}) of test particles can be
derived from $\delta \ ^{p}S=0,$ where
\begin{equation*}
\ ^{p}S=\int \ ^{p}Lds,\ \ ^{p}L=e^{q}\sqrt{|\mathbf{g}_{\alpha \beta
}v^{\alpha }v^{\beta }|}.
\end{equation*}%
and limit of weak gravitational fields of $ds$ is characterized by  $%
ds\approx (1+2\varphi -\overrightarrow{v}^{2})^{1/2}\approx (1+\varphi -%
\overrightarrow{v}^{2}/2)dt$. In above formula, $\varphi $ is the Newton
potential and $\overrightarrow{v} $ is the velocity of the fluid in the 3--d
space.

The solution (\ref{sol5e}) can approximated  $e^{q}\approx 1+U(\rho ,F)$,
for
\begin{equation*}
U(\rho ,F)=w[\ ^{1}a+\ ^{2}a(F)-(\ ^{1}b+\ ^{2}b(F))\rho _{0}]^{-1}\ln
\{C\rho /[\ ^{1}a+\ ^{2}a(F)+(\ ^{1}b+\ ^{2}b(F))(\rho -\rho _{0})]\}
\end{equation*}%
and the variation of respective action is  $\delta \ ^{p}S=\delta \left[
1+U(\rho ,F)+\varphi -\overrightarrow{v}^{2}/2\right] dt=0$. This allows us
to compute the 3--d acceleration of the particle using the 3--d gradient ${%
grad}$ (for nontrivial $F,$ we should perform a N--adapted calculus of this
gradient),
\begin{equation*}
\ ^{tot}\overrightarrow{a}=-{grad\ }[\varphi +U(\rho ,F)]=\ ^{s}%
\overrightarrow{a}+\ ^{p}\overrightarrow{a}+\ ^{N}\overrightarrow{a}+\ ^{E}%
\overrightarrow{a},
\end{equation*}%
where $\ ^{s}\overrightarrow{a}=-{grad\ }\varphi $
\begin{eqnarray}
\ ^{s}\overrightarrow{a} &=&-{grad\ }\varphi ,%
\mbox{\  the Newtonian
gravitational accelleration };  \notag \\
\ ^{p}\overrightarrow{a}(\rho ,p,F) &=&-\frac{C}{\ ^{1}a+\ ^{2}a(F)-(\
^{1}b+\ ^{2}b(F))\rho _{0}}\frac{1}{\rho }\ {grad\ }p,  \notag \\
&&\mbox{\  the
hydrodinamical accelleration };  \notag \\
\ ^{N}\overrightarrow{a}(...,F) &\approx &\mbox{\  N--connection terms  };
\label{accf} \\
\ ^{E}\overrightarrow{a}(\rho ,p,F) &=&\frac{\ ^{1}b+\ ^{2}b(F)}{1+\ ^{1}a+\
^{2}a(F)-(\ ^{1}b+\ ^{2}b(F))\rho _{0}}\times \frac{{grad\ }p}{1+\ ^{1}a+\
^{2}a(F)+(\ ^{1}b+\ ^{2}b(F))(\rho _{0}-\rho )},  \notag
\end{eqnarray}%
where the last term $\ ^{E}\overrightarrow{a}$ is a supplementary
acceleration induced from modifications of the action and (Finsler)
anisotropies of gravitational field. \ Such a term, $\ ^{E}\overrightarrow{a}%
\simeq \ ^{E}a(\rho ,p),$ was used in Section V.C \ of Ref. \cite{harko1}
for computing possible modification of perihelion procession of Mercury
\begin{equation*}
\bigtriangleup \varphi =\frac{6\pi GM_{\odot }}{a(1-\epsilon ^{2})}+\frac{%
2\pi a^{2}\sqrt{1-\epsilon ^{2}}}{GM_{\odot }}\ (\ ^{E}a),
\end{equation*}%
where $\epsilon $ is the eccentricity of orbit, $a$ is the distance between
Mercury and Sun, $M_{\odot }$ is the Sun's mass. The observational data
constrain for modifications resulting from a $f(R,\rho )$ theory where
computed $\ ^{E}a\leq 1.28\times 10^{-9}$ cm/s$^{2}.$ The terms $\ ^{p}%
\overrightarrow{a},\ ^{N}\overrightarrow{a},\ ^{E}\overrightarrow{a}$ in (%
\ref{accf}) depend anisotropically on $F.$ So, we have to take into
consideration such terms when the anisotropic effects and constraints for
the $f(R,\rho ,F)$ models are computed.

\section{Decoupling \& Integrability of $f(R,T,F)$ Gravity}

\label{sdif}

The field equations in modified gravity theories are very "sophisticated"
systems of nonlinear PDE. Surprisingly, it is possible to decouple and
integrate such PDE in general forms using the anholonomic deformation method
\cite{veyms,vofdc} (see also references therein). In this section, we
reformulate the method for constructing generic off--diagonal solutions for
the $f(R,T,F)$ gravity, with trivial or nontrivial contributions from a
Finsler generating function $F,$ when $\mathbf{g}_{\alpha \beta }(u^{\gamma
})=\mathbf{\tilde{g}}_{\alpha \beta }(u^{\gamma },y^{\alpha }(u^{\mu }))$ (%
\ref{osculef}). Finally, we shall provide examples of solutions for
ellipsoid and solitonic configurations.

\subsection{The anholonomic deformation method for modified gravity}

For simplicity, we shall prove integrability of\ the system (\ref{eeff1})
for $\mathbf{T}_{\beta \gamma }=0,$%
\begin{eqnarray}
\mathbf{\mathbf{R}}_{\ \gamma }^{\beta }&=&\frac{1}{2}\Upsilon (x^{i})\delta
_{\ \gamma }^{\beta },  \label{einstdeq} \\
\mbox{ where \qquad } \frac{1}{2}\Upsilon (x^{i})&:=&\tilde{\Lambda}%
(x^{i})+\ [\ ^{2}f+2\partial _{T}(\ ^{2}f)],  \label{source3}
\end{eqnarray}%
for $[\ ^{2}f+2\partial _{T}(\ ^{2}f)]_{\mid T=0}\ $\ being a nontrivial
function on $x^{i}$ and an effective anisotropically polarized cosmological
"constant" $\tilde{\Lambda}(x^{i})=\frac{1}{2}[\ ^{h}R(x^{i})+\Lambda
(x^{i})].$ The source $\Upsilon (x^{i})$ contains information on possible
contributions from a Finsler generating function $F$ and modifications by $\
^{2}f(\mathbf{T}).$ The solutions of this system of nonlinear PDE define
nonholonomic Einstein manifolds.

\subsubsection{Decoupling of field equations}

The decoupling property can be proven for metrics with one Killing symmetry
on $\partial /\partial y^{4},$ with local coordinates $u^{\alpha
}=(x^{1},x^{2},y^{3},y^{4}),$ when
\begin{equation}
\mathbf{g}=\epsilon _{i}e^{\psi (x^{k})}dx^{i}\otimes
dx^{j}+h_{3}(x^{k},y^{3})\mathbf{e}^{3}\otimes \mathbf{e}%
^{3}+h_{4}(x^{k},y^{3})\mathbf{e}^{4}\otimes \mathbf{e}^{4},  \label{ans1}
\end{equation}%
for $\mathbf{e}^{3}=dy^{3}+w_{i}(x^{k},y^{3})dx^{i},\ \mathbf{e}%
^{4}=dy^{4}+n_{i}(x^{k},y^{3})dx^{i}$. Such metrics are of type (\ref{ddif})
up to frame transforms, $\epsilon _{i}=\pm 1$ depending on a chosen
signature for the spacetime metric. We shall use brief denotations of
partial derivatives, for instance, $s^{\bullet }=\partial /\partial
x^{1},s^{\prime }=\partial /\partial x^{2}$ and $s^{\ast }=\partial
/\partial y^{3}$ and construct exact solutions in such N--adapted frames
when $h_{4}^{\ast }\neq 0$ and $\Upsilon (x^{i})\neq 0.$

\paragraph{Gravitational field equations for $\mathbf{D:}$}

For ansatz (\ref{ans1}), the nonlinear PDEs (\ref{einstdeq}) are equivalent
to
\begin{eqnarray}
\epsilon _{1}\psi ^{\bullet \bullet }+\epsilon _{2}\psi ^{\prime \prime }
&=&\Upsilon ,  \label{eq1} \\
\phi ^{\ast }(\ln |h_{4}|)^{\ast } &=&\Upsilon h_{3},  \label{eq2} \\
\beta w_{i}+\alpha _{i} &=&0,  \label{eq3} \\
n_{i}^{\ast \ast }+\gamma n_{i}^{\ast } &=&0,  \label{eq4}
\end{eqnarray}%
where the coefficients
\begin{equation}
\gamma = (\ln |h_{4}|^{3/2}-\ln |h_{3}|)^{\ast },\ \alpha _{i} = h_{4}^{\ast
}\partial _{i}\phi \mbox{ and }\beta =h_{4}^{\ast }\phi ^{\ast },
\label{coeff}
\end{equation}%
are determined by $h_{3}$ and $h_{4}$ via
\begin{equation}
\phi =\ln |2(\ln \sqrt{|h_{4}|})^{\ast }|-\ln \sqrt{|h_{3}|},  \label{genf}
\end{equation}
see detailed computations of the N--adapted coefficients for the Ricci and
Einstein tensors in \cite{veyms,vofdc}.

The above system of equations (\ref{eq1})-(\ref{eq4}) reflects a very
important decoupling property of the Einstein equations for certain classes
of metric compatible linear connections and with respect to N--adapted frame
(in this section, we consider for simplicity only metrics with one Killing
symmetry):

\begin{enumerate}
\item \textbf{\ }Depending on signature, the equation (\ref{eq1}) is a 2--d
D'Alambert, or Laplace, equation which can be integrated for arbitrary
source $\Upsilon (x^{k}).$

\item The system of two equations (\ref{eq2}) and (\ref{genf}) is for three
unknown functions $h_{3}(x^{k},y^{3}),h_{4}(x^{k},y^{3})$ and $\phi
(x^{k},y^{3})$ if a source $\Upsilon (x^{k})$ is prescribed. It contains
only partial derivatives $\ast =\partial /\partial y^{3}.$ We can integrate
in general form and define certain functionals $h_{3}[\phi ]$ and $%
h_{4}[\phi ]$ for any prescribed generating function $\phi ,\phi ^{\ast
}\neq 0,$ and integration functions and parameters, see below formulas (\ref%
{gensol1}).

\item The equations (\ref{eq3}) and (\ref{eq4}) are respectively algebraic
ones (for $w_{i}$) and contains only first and second derivatives on $%
\partial /\partial y^{3}$ of $n_{i}.$ For any defined $h_{3}$ and $h_{4},$
we can compute the coefficients $\alpha _{i},\beta $ and $\gamma $ following
formulas (\ref{coeff}) and integrate all equations for $N$--coefficients in
general form.
\end{enumerate}

We conclude that with respect to N--adapted frames (\ref{dder}) and (\ref%
{ddif}) determined by $N_{i}^{a}=(w_{i},n_{i})$ the modified Einstein
equations (\ref{einstdeq}) decouple into PDE with \ derivatives of 2d and
1st order, and algebraic equations for corresponding coefficients of metric.%
\footnote{%
Via conformal and frame transforms and introducing additional multiples, we
can prove decoupling properties for various classes of metrics depending on $%
y^{4},$ i.e. on all coordinated on a manifold $\mathbf{V}$ of finite
dimension. For simplicity, we omit such constructions in this work.} This
property can be proven in explicit form for $\mathbf{D}$ and contain
additional information on $f(R,T,F)$ modifications of gravity via source $%
\Upsilon (x^{k})$ and induced torsion of $\mathbf{D.}$

In this work we construct solutions with nonzero $\phi ^{\ast }$ and $%
\Upsilon (x^{k})$ because we are interested to investigate possible
nontrivial contributions from modified gravity via nontrivial sources $%
\Upsilon (x^{k}). $ Vacuum configurations $\phi ^{\ast }=0$ and $\Upsilon
(x^{k})=0$ can be studies by similar methods, see \cite{veyms,vofdc}.

\paragraph{Constraints for the Levi--Civita connection $\mathbf{\protect%
\nabla :}$}

For ansatz of type (\ref{ans1}), the zero--torsion constraints (\ref%
{lcconstr}) can be satisfied if
\begin{equation}
w_{i}^{\ast }=\mathbf{e}_{i}\ln |h_{4}|,\partial _{i}w_{j}=\partial
_{j}w_{i},n_{i}^{\ast }=0.  \label{lcconstr1}
\end{equation}%
The first condition "brock" the decoupling property of the system (\ref{eq1}%
)-(\ref{eq4}). The constraints of vanishing the torsion for $\nabla $ relate
additionally the N--coefficients with $h_{4}.$ Nevertheless, we can solve
such conditions in explicit form via additional frame and coordinate
transforms and/or re--parametrization of generating functions etc. For
instance, we may fix any convenient value for $\ln |h_{4}|$ and use it as a
generating function in the system (\ref{eq2}) and (\ref{genf}) in order to
define $h_{3}$ and $\phi .$ Next step will consist in determining $w_{i}$
from algebraic equations (\ref{eq3}). The equations (\ref{eq4}) became
trivial for $n_{i}^{\ast }=0$ which allows us to introduce any $%
n_{i}(x^{k}),\partial _{k}n_{i}=\partial _{i}n_{k}$, in the off--diagonal
metric ansatz.

In general, we can consider (\ref{lcconstr1}) as a class of nonholonomic
constraints on integral varieties of solutions for $\mathbf{D}$ which
results in subvarieties with torsionless configurations for $\nabla .$

\subsubsection{General solutions for modified field equations}

The $h$--metric is given by $\epsilon _{i}e^{\psi (x^{k})}dx^{i}\otimes
dx^{j}$, where $\psi (x^{k})$ is a solution of (\ref{eq1}) considered as a
2--d d'Alambert/ Laplace equation (\ref{eq1}). It depends on $f(R,T,F)$ via
source $\Upsilon (x^{k})$ (\ref{source3}).

As a second step, we integrate in general form the system (\ref{eq2}) and (%
\ref{genf}), for $\phi ^{\ast }\neq 0.$ Defining $A:=(\ln |h_{4}|)^{\ast }$
and $B=\sqrt{|h_{3}|},$ we re--write such equations in the form%
\begin{equation}
\phi ^{\ast }A=\Upsilon B^{2},\ Be^{\phi }=2A.  \label{aux6e}
\end{equation}%
Considering $B\neq 0,$ we obtain $B=(e^{\phi })^{\ast }/2\Upsilon $ as a
solution of a system of quadratic algebraic equations. This formula can
integrated on $dy^{3}$ which allows us to find $\sqrt{|h_{3}(x^{k},y^{3})|}=%
\sqrt{|\ ^{0}h_{3}(x^{k})|}+\partial _{3}e^{\phi (x^{k},y^{3})}/2\Upsilon
(x^{k})$. We can write
\begin{equation}
h_{3}=\ ^{0}h_{3}(1+(e^{\phi })^{\ast }/2\Upsilon \sqrt{|\ ^{0}h_{3}|})^{2}
\label{solh3}
\end{equation}%
if the local signature of the term $\ ^{0}h_{3}$ is the same as $h_{3}.$
Introducing the value $h_{3}$ in (\ref{aux6e}) and integrating on $y^{3},$
we find
\begin{equation*}
h_{4}=\ ^{0}h_{4}\exp [(8\Upsilon )^{-1}\ e^{2\phi }],
\end{equation*}%
where $\ ^{0}h_{4}=$ $\ ^{0}h_{4}(x^{k})$ is an integration function.

The N--connection coefficients can be found from (\ref{eq3}), $%
w_{i}=-\partial _{i}\phi /\phi ^{\ast },$ and integrating two times on $%
y^{3} $ in (\ref{eq4}),%
\begin{equation}
n_{k}=\ ^{1}n_{k}+\ ^{2}n_{k}\int dy^{3}h_{3}/(\sqrt{|h_{4}|})^{3},
\label{ncoeff}
\end{equation}%
where $\ ^{1}n_{k}(x^{i})$ and $\ ^{2}n_{k}(x^{i})$ are integration
functions.

Putting all terms together in (\ref{ans1}), we obtain a formal general
solution of the gravitational field equations (\ref{einstdeq}) in $f(R,T,F)$
gravity via quadratic element
\begin{eqnarray}
ds^{2} &=&\epsilon _{i}e^{\psi \lbrack \Upsilon ]}(dx^{i})^{2}+\ ^{0}h_{3}(1+%
\frac{(e^{\phi })^{\ast }}{2\Upsilon \sqrt{|\ ^{0}h_{3}|}})^{2}\ [dy^{3}-%
\frac{\partial _{i}\phi }{\phi ^{\ast }}dx^{i}]^{2}  \label{gensol1} \\
&&+\ ^{0}h_{4}\exp [(8\Upsilon )^{-1}e^{2\phi }][dy^{4}+(\ ^{1}n_{k}+\
^{2}n_{k}\int dy^{3}\frac{h_{3}}{(\sqrt{|h_{4}|})^{3}})dx^{i}]^{2}.  \notag
\end{eqnarray}
Such solutions depend on generating functions $\phi (x^{i},y^{3})$ and $\psi
\lbrack \Upsilon (x^{k})]$ and on integration functions $\ ^{0}h_{3}(x^{k}),$
\newline
$\ ^{0}h_{4}(x^{k}),$ $\ ^{1}n_{k}(x^{k}),\ ^{2}n_{k}(x^{k})$ as we
described in above formulas. This class of modified spacetimes are
characterized by nontrivial torsion with coefficients computed following
formulas (\ref{dtors}) using only the coefficients of metric (and respective
N--connection). In general, we can introduce additional parameters and
derive new symmetries because of existing Killing symmetry, see \cite%
{vrflg,veyms,vofdc}. We can also consider that the class of solutions (\ref%
{gensol1}) is for the $f(R,T)$ gravity when certain Finsler like variables
where introduced in order to be able to decouple the field equations and get
very general classes of solutions as effective 4-d nonholonomic Einstein
equations.

Constraining additionally the class of generating and integration functions
in (\ref{gensol1}), we construct exact solutions for the Levi--Civita
connection $\nabla .$ We have to consider solutions with $\
^{2}n_{k}=0,\partial _{i}(\ ^{1}n_{k})=\partial _{k}(\ ^{1}n_{i})$ and $%
w_{i}=-\partial _{i}\phi /\phi ^{\ast }$ and $h_{4}$ are subjected to
conditions (\ref{lcconstr1}). Even for solutions with $\nabla ,$ we get only
effective Einstein spaces with locally anisotropic polarizations. This is
because the source $\Upsilon (x^{k})$ (\ref{source3}) determines the
diagonal coefficients of (\ref{gensol1}), with respect to N--adapted frames.
This results, in general, in effective polarization of the gravitational
constant as in (\ref{effgc1}). We can generate solutions for Einstein
manifolds if we fix $\Upsilon (x^{k})=const.$

Finally, we note that metrics of type (\ref{gensol1}) are generic
off--diagonal, i.e. we can not diagonalize such solutions via coordinate
transform. This follows from the fact that the anholonomy coefficients, see
formulas (\ref{anhrel}), are not zero for arbitrary generating and
integration functions.

\subsection{Examples of exact solutions}

Quadratic elements (\ref{gensol1}) parameterize formal integrals of a system
of nonlinear PDE for modified gravity. It may describe certain physical real
situations if the generating and integration functions and parameters are
subjected to realistic boundary/ asymptotic conditions with associated
symmetries and conservation laws.

In spherical coordinates $u^{\alpha }=(x^{1}=r,x^{2}=\theta ,y^{3}=\varphi
,y^{4}=t),$ a diagonal metric%
\begin{equation}
\ ^{\circ }\mathbf{g}=\underline{q}^{-1}(r)dr\otimes dr+r^{2}d\theta \otimes
d\theta +r^{2}\sin ^{2}\theta d\varphi \otimes d\varphi -\underline{q}%
(r)dt\otimes dt,  \label{dsbh}
\end{equation}
defines an empty de Sitter space if $\underline{q}(r)=1-$ $2\frac{m(r)}{r}%
-\Lambda \frac{r^{2}}{3},$ where $\Lambda $ is a cosmological constant. The
total mass--energy within the radius $r$ is defined by a function $m(r).$
For $m(r)=0$ we obtain an empty space with a cosmological horizon at $%
r=r_{c}=\sqrt{3/\Lambda }.$ If $m(r)=m_{0}=const$ and $\Lambda =0,$ we get
the Schwarzschild solution. The metric (\ref{dsbh}) is an example of
diagonal solution of (\ref{einstdeq}) in GR, when $f(R,T,F)=R,$ $N_{i}^{a}=0,%
\mathbf{D}=\nabla $ and $\frac{1}{2}\Upsilon (x^{i})=\Lambda =const,$ see
source (\ref{source3}).

In this section, we analyze two classes of solutions related to possible $%
f(R,T,F)$ modifications of GR. In the first case, we construct metrics for
possible off--diagonal deformations and polarizations of coefficients of de
Sitter black holes resulting in ellipsoidal configurations. In the second
case, the de Sitter black holes are embedded self--consistently into certain
solitonic background configurations.

\subsubsection{Ellipsoid configurations in $f(R,T,F)$ gravity}

The generic off--diagonal ansatz is chosen
\begin{eqnarray}
\ ds^{2} &=&e^{\psi (\xi ,\vartheta )}\ (d\xi ^{2}+\ d\vartheta
^{2})+h_{3}(\xi ,\vartheta ,\varphi )\ (\mathbf{e}_{\varphi })^{2}+h_{4}(\xi
,\vartheta ,\varphi )\ (\mathbf{e}_{t})^{2},  \label{anselconf} \\
~\mathbf{e}_{\varphi } &=&d\varphi +w_{1}\left( \xi ,\vartheta ,\varphi
\right) d\xi +w_{2}\left( \xi ,\vartheta ,\varphi \right) d\vartheta ,\
\mathbf{e}_{t}=dt+n_{1}\left( \xi ,\vartheta ,\varphi \right) d\xi
+n_{2}\left( \xi ,\vartheta ,\varphi \right) d\vartheta ,  \notag
\end{eqnarray}
for $h_{3}=\eta _{3}(\xi ,\theta ,\varphi )r^{2}(\xi )\sin ^{2}\theta ,\
h_{4}=\eta _{4}(\xi ,\theta ,\varphi )\ \varpi ^{2}(\xi )$, local
coordinates $x^{1}=\xi ,x^{2}=\vartheta =r(\xi )\theta ,y^{3}=\varphi
,y^{4}=t,$ with $\xi =\int dr/\left\vert \underline{q}(r)\right\vert ^{\frac{%
1}{2}}.$ We get a diagonal configuration if $w_{i}=0,n_{i}=0,\eta
_{3}=1,\eta _{4}=1$ and $\psi =0,$
\begin{equation}
\ ^{\circ }\mathbf{g}=d\xi \otimes d\xi +r^{2}(\xi )\ d\theta \otimes
d\theta +r^{2}(\xi )\sin ^{2}\theta \ d\varphi \otimes d\varphi -\underline{q%
}(\xi )\ dt\otimes \ dt,  \label{dsbha}
\end{equation}%
with coefficients $\check{g}_{1}=1,\ \check{g}_{2}=r^{2}(\xi ),\ \check{h}%
_{3}=r^{2}(\xi )\sin ^{2}\theta ,\ \check{h}_{4}=-q(\xi )$. In variables $%
\left( r,\theta ,\varphi \right) ,$ the metric (\ref{dsbha}) is equivalent
to (\ref{dsbh}).

The ansatz (\ref{anselconf}) is an example of solutions of type (\ref%
{gensol1}) if the coefficients are generated following similar methods
taking $\ ^{0}h_{3}=\ \check{h}_{3}$ and $\ ^{0}h_{4}=\ \check{h}_{4},$

The coefficients \ of this metric determine exact solutions if {\small
\begin{eqnarray}
&&\psi ^{\bullet \bullet }(\xi ,\vartheta )+\psi ^{^{\prime \prime }}(\xi
,\vartheta )=\Upsilon (\xi ,\vartheta );  \label{anhsol2} \\
h_{3} &=&\ \check{h}_{3}(1+\partial _{\varphi }(e^{\phi })/2\Upsilon \sqrt{|%
\check{h}_{3}|})^{2},\ h_{4}=\ \check{h}_{4}\exp [(8\Upsilon )^{-1}e^{2\phi
}];  \notag \\
w_{i} &=&-\partial _{i}\phi /\phi ^{\ast };\ n_{i}=\ ^{1}n_{i}(\xi
,\vartheta )+\ ^{2}n_{i}(\xi ,\vartheta )\int d\varphi h_{3}/(\sqrt{|h_{4}|}%
)^{3},  \notag
\end{eqnarray}%
} for any nonzero $h_{a}$ and $h_{a}^{\ast },$ and (integrating) functions $%
^{1}n_{i}(\xi ,\vartheta ),\ ^{2}n_{i}(\xi ,\vartheta )$ and generating
function $\phi (\xi ,\vartheta ,\varphi ).$

For nonholonomic ellipsoid de Sitter configurations (for simplicity, we
consider rotoid configurations with small eccentricity $\varepsilon $), we
parameterize
\begin{eqnarray}
~_{\lambda }^{rot}\mathbf{g} &=&e^{\psi (\xi ,\vartheta )}\ (d\xi ^{2}+\
d\vartheta ^{2})+r^{2}(\xi )\sin ^{2}\theta \ \left( 1+\partial _{\varphi
}e^{\phi }/2\Upsilon \sqrt{|\underline{q}|}\right) ^{2}\ ~\mathbf{e}%
_{\varphi }\otimes \mathbf{e}_{\varphi }-\left[ \underline{q}(\xi
)+\varepsilon \zeta (\xi ,\vartheta ,\varphi )\right] \ \mathbf{e}%
_{t}\otimes \mathbf{e}_{t},  \notag \\
~\mathbf{e}_{\varphi } &=&d\varphi -\frac{\partial _{\xi }\phi }{\partial
_{\varphi }\phi }d\xi -\frac{\partial _{\vartheta }\phi }{\partial _{\varphi
}\phi }d\vartheta ,\ \mathbf{e}_{t}=dt+n_{1}(\xi ,\vartheta ,\varphi )d\xi
+n_{2}(\xi ,\vartheta ,\varphi )d\vartheta ,  \label{soladel}
\end{eqnarray}%
where $n_{i}$ (\ref{ncoeff}) are computed in some forms $n_{i}\sim
\varepsilon ...\ $ for corresponding coordinates and values $h_{3}$ and $%
h_{4}.$ The function
\begin{equation}
\zeta =\ \underline{\zeta }(\xi ,\vartheta )\sin (\omega _{0}\varphi
+\varphi _{0}),  \label{rhorot}
\end{equation}%
for some constant parameters $\omega _{0}$ and $\varphi _{0},$ we can state $%
\underline{\zeta }(\xi ,\vartheta )\simeq \underline{\zeta }=const,$ is
chosen to generate an anisotro\-pic rotoid configuration for the smaller
\textquotedblleft horizon\textquotedblright\ (when the term before $\mathbf{e%
}_{t}\otimes \mathbf{e}_{t}$ became$\ h_{4}=0),\ $%
\begin{equation*}
r_{+}\simeq 2\ m_{0}/\left( 1+\varepsilon \underline{\zeta }\sin (\omega
_{0}\varphi +\varphi _{0})\right) ,
\end{equation*}%
where $\varepsilon $ is the eccentricity. The generating function $\
_{\shortmid }\phi (\xi ,\tilde{\vartheta},\varphi )$ contained in (\ref%
{anhsol2}) is related to $\zeta (\xi ,\tilde{\vartheta},\varphi )$ via
formula $e^{2\phi }=8\Upsilon \ln |1-\varepsilon \zeta /\underline{q}(\xi
)|, $ for which a rotoid configuration (\ref{rhorot}) can be fixed. We
construct rotoid deformations of the de Sitter black hole metric (\ref{dsbh}%
) if introduce the function $\zeta $ (\ref{rhorot}) into the last formula
and define a generating function/functional $\phi =\phi (\underline{q}%
,\Upsilon ,\zeta ,F).$ In general, such off--diagonal deformations do not
result in other classes of black hole solutions. If we consider small
deformations on parameter $\varepsilon $ for which
\begin{equation*}
h_{3}=\ \check{h}_{3}(1+\varepsilon \chi _{3}),h_{4}=\ \check{h}%
_{4}(1+\varepsilon \chi _{4}),w_{i}\sim \varepsilon \check{w}_{i},n_{i}\sim
\varepsilon \check{n},
\end{equation*}%
metrics of type (\ref{soladel}) describe stationary black ellipsoid
solutions with coefficients computed with respect to N--adapted frames, see
discussion and references in \cite{vrflg,veyms}. If we restrict the integral
variety of such solutions to satisfy the conditions (\ref{lcconstr1}), we
generate exact off--diagonal solutions for the Levi--Civita connection $%
\nabla .$

\subsubsection{Black holes \& locally anisotropic solitonic backgrounds}

Another example of off--diagonal solutions with local anisotropies in
modified gravity can be constructed as a nonlinear superposition of the de
Sitter black hole solution and solitonic waves. We consider a
non--stationary ansatz
\begin{eqnarray}
ds^{2} &=&e^{\psi (\xi ,\vartheta )}[d\xi ^{2}+d\vartheta ^{2}]-\underline{q}%
(\xi )(1+\frac{\partial _{t}e^{\phi (\xi ,\vartheta ,t)}}{2\Upsilon (\xi
,\vartheta )\sqrt{|\underline{q}(\xi )|}})^{2}[dt-\frac{\partial _{\xi }\phi
}{\partial _{t}\phi }d\xi -\frac{\partial _{\vartheta }\phi }{\partial
_{t}\phi }d\vartheta ]^{2}  \notag \\
&&\ +r^{2}(\xi )\sin ^{2}\vartheta \exp [(8\Upsilon (\xi ,\vartheta
))^{-1}e^{2\phi (\xi ,\vartheta ,t)}][d\varphi +(\ ^{1}n_{1}(\xi ,\vartheta
)+\ ^{2}n_{1}(\xi ,\vartheta )\int dt\frac{h_{3}(\xi ,\vartheta ,t)}{(\sqrt{%
|h_{4}(\xi ,\vartheta ,t)|})^{3}})d\xi  \notag \\
&&+(\ ^{1}n_{2}(\xi ,\vartheta )+\ ^{2}n_{2}(\xi ,\vartheta )\int dt\frac{%
h_{3}(\xi ,\vartheta ,t)}{(\sqrt{|h_{4}(\xi ,\vartheta ,t)|})^{3}}%
)d\vartheta ]^{2},  \label{anssol1}
\end{eqnarray}
for local coordinates $x^{1}=\xi ,x^{2}=\vartheta ,y^{3}=t,y^{4}=\varphi $
and $\underline{q}(\xi )=\underline{q}(r(\xi ))$ computed as in (\ref{dsbh}).

\paragraph{Solitonic backgrounds with radial Burgers equation:}

We take $\phi (\xi ,\vartheta ,t)=\eta (\xi ,\vartheta ,t),$ when $y^{3}=t$
is a time like coordinate, as a solution of KdP equation \cite%
{kadom,vsoliton},
\begin{equation}
\pm \eta ^{^{\prime \prime }}+(\partial _{t}\eta +\eta \ \eta ^{\bullet
}+\epsilon \eta ^{\bullet \bullet \bullet })^{\bullet }=0,  \label{kdp1}
\end{equation}%
with dispersion $\epsilon $ and possible dependencies on a set of parameters
$\theta .$ It is supposed that in the dispersionless limit $\epsilon
\rightarrow 0$ the solutions are independent on $x^{2}$ and determined by
Burgers' equation $\partial _{t}\eta +\eta \ \eta ^{\bullet }=0.$
Introducing generating functions $\phi $ determined by solutions of such
3--d solitonic equations in (\ref{anssol1}), we generate solitonic
nonholonomic deformations of the de Sitter black hole solutions. In general,
the new off--diagonal solutions do not have black hole properties. We can
consider other types of solitonic solutions. Such configurations always
define exact solutions of gravitational field equations (\ref{einstdeq}) for
$\mathbf{D}.$ Constraining the solitonic integral varieties via conditions (%
\ref{lcconstr1}), we generate solutions for the Levi--Civita connection $%
\nabla .$

\paragraph{Solitonic backgrounds with angular Burgers equation:}

In this case $\phi =\widehat{\eta }(\xi ,\vartheta ,t)$ is a solution of KdP
equation
\begin{equation}
\pm \widehat{\eta }^{\bullet \bullet }+(\partial _{t}\widehat{\eta }+%
\widehat{\eta }\ \widehat{\eta }^{\prime }+\epsilon \widehat{\eta }^{\prime
\prime \prime })^{\prime }=0.  \label{kdp3a}
\end{equation}%
In the dispersionless limit $\epsilon \rightarrow 0$ the solutions are
independent on $x^{1}=\xi $ and determined by Burgers' equation $\partial
_{t}\widehat{\eta }+\widehat{\eta }\ \widehat{\eta }^{\prime }=0.$
Introducing $\phi =\widehat{\eta }$ in (\ref{anssol1}), we generate
solutions of (\ref{einstdeq}) with angular anisotropy. \ For small values of
\ $\widehat{\eta },$ we can models 3--d solitonic polarizations of the de
Sitter black holes.

Finally, we note that nonholonomic constraints and off--diagonal
interactions with terms induced from modified and/or locally anisotropic
gravity (for instance, via solitonic waves) may preserve the black hole
character of certain classes or "disperse" them into effective nonlinear
vacuum configurations with polarized cosmological constants.

\section{Discussion and Conclusions}

\label{sdc}

In this work, we have merely presented a flavour of a geometric formalism in
modified $f(R,T)$ theories starting from principles of general covariance
and relativity for nonholonomic deformations of fundamental
geometric/physical objects and field equations on Lorentz manifolds and
their tangent bundles. The goal was to prove a general decoupling property
and further integrability in very general forms of gravitational field
equations. Surprisingly, such systems of nonlinear PDE can be solved in
general forms with respect to certain classes of nonholonomic frames for
(auxiliary) Finsler type connections. Imposing nonholonomic constraints on
certain classes of generic off--diagonal integral varieties, the solutions
can be transformed into configurations for the torsionless Levi--Civita
connection.

There are two very different approaches to connect Finsler like geometries
to Einstein gravity and modifications:

\begin{enumerate}
\item The first one is to introduce on (pseudo) Riemannian / Lorentz
manifolds a non--integrable (nonholonomic) 2+2 splitting with a conventional
fibred structure. Such constructions are very similar to those in
Finsler--Cartan geometry with metric compatible connections. A reason to
introduce Finsler like variables in GR is that we can elaborate a geometric
method of integrating the gravitational field equations. Here we note that
it is possible to elaborate also certain new geometric methods of
quantization of gravity theories using almost K\"{a}hler -- Finsler
variables (following respective gauge like, A--brane and deformation
quantization formalisms). In such cases, a Finsler geometry is modelled via
nonholonomic distributions as an auxiliary tool and does not change the
paradigm of an originally considered Einstein or modified gravity theory.

\item The second approach is related to more fundamental locally anisotropic
modifications of the concept of spacetime and gravitational interactions.
They can be motivated by certain theoretical arguments, for instance, in
quantum gravity, modified dispersion relations, and theories with possible
violations of local Lorentz symmetry etc. Roughly speaking, a very general
class of modified theories of gravity have to be elaborated on tangent
bundles of Lorentz manifolds when certain geometric/physical principles are
used for extensions of GR to Einstein--Finsler type gravity theories. Such
models may play a physically important role because they accept an axiomatic
very similar to that for the Einstein gravity, such theories can integrated
in very general form and there are known well defined methods of
quantization of such models. Certain anisotropic solutions from Finsler
gravity seem to play an important role in explaining locally anisotropic
effects in modern cosmology. On tangent Lorentz manifolds, the gravity
models are for an extra dimensional spacetimes (eight dimensions) which can
be reduced to "effective" 4-d metrics via osculating procedure.
\end{enumerate}

In another turn, the late--time acceleration of the Universe and dark
energy/ matter effects are intensively studied in the framework of theories
with modifications of Lagrange density, $R\to f(R,T,...)$. Various such
models with exotic anisotropic states of matter, modified gravitational and
matter field interactions, torsion etc were elaborated. The corresponding
field equations are very sophisticated nonlinear PDE which request advanced
analytic and numerical methods for constructing solutions and analysis of
possible physical implications. One of the main goals of this work is to
extend the anholonomic deformation method of constructing exact solutions in
Einstein and/or Finsler gravity theories in such a form which would allow to
integrate in very general forms certain classes of $f(R,T,...)$ gravity
models. From a "modest" pragmatic point of view, a Finsler generating
function $F$ is just a formally prescribed nonholonomic distribution on
various spacetime models which allows us to decouple and solve physically
important field equations. But the method may be also formulated to encode
possible modifications from locally anisotropic gravity models on tangent
bundles. Conventionally, various geometric and physical assumptions for
modified gravity theories are denoted as $f(R,T,F)$. We proved that via
nonholonomic frame transforms and deformations we can model a $f(R,T,F)$
theory as a $f(R,T)$ one, or as an effective Einstein, or Finsler, theory.

In the present work, we investigated generalized gravity theories with
arbitrary coupling (including anisotropi\-es, parametric dependencies and
nonholonomic constraints) between matter and geometry. Using geometric and
variational methods, all adapted to possible nonlinear connection
structures, we derived the gravitational field equations. We considered
several important particular cases that may present interest in modern
cosmology and astrophysics. We concluded that off--diagonal terms of
metrics, modified matter and time dependent terms in generalized Einstein
equations play the role of effective cosmological constant, exotic matter,
effective torsion fields, anholonomic frame effects etc. For well defined
geometric and physical conditions, such effects can be modified by
nonholonomic distributions on Einstein manifolds and their tangent bundles.

We studied the modified equations of motion of test particles in modified $%
f(R,T,F)$ theories and evaluated possible contributions of effective
extra--forces with locally anisotropic terms. There were formulated the
nonholonomically modified Raychaudhury equations on tangent Lorentz bundle.
Certain Newton limits with corresponding corrections were computed. Using
perihelion procession, an upper limit to extra--acceleration and
anisotropies was obtained. As an explicit example, we took the de Sitter
black hole metric and deformed it into new classes of exact solutions with
rotoid symmetry. The effective source (an anisotropically polarized
cosmological constant) contains contributions from possible modifications of
gravitational actions and/or from Finsler like anisotropies. We provided
some examples when black hole solutions are modified ("dissipated") into
off--diagonal vacuum configurations with complex locally anisotropic
structure and effective cosmological constant.

We conclude that the predictions of $f(R,T,F)$ theories could be very
different from those in $f(R,T)$, $f(R)$ and/or GR theories. Generalized
principles of covariance and relativity may state certain conditions of
equivalence and mutual transforms of various models. The study of these
theoretical issues and related phenomena may provide some specific effects
which may help to distinguish different gravitational models. In our
forthcoming work, we shall explore in more details such theories by
elaborating certain models of modified/anisotropic cosmological evolution
and possible dark energy/matter effects.

\vskip5pt \textbf{Acknowledgements: } SV research is partially supported by
the Program IDEI, PN-II-ID-PCE-2011-3-0256 and a visiting research program
at CERN.

\end{document}